\RequirePackage{fix-cm}
\documentclass[natbib,smallextended]{svjour3}       
\smartqed  
\usepackage{graphicx,color}
 \journalname{my journal}
\usepackage{epsfig}
\usepackage{amsmath}
\usepackage{enumerate}
\usepackage{hyperref}
\usepackage{ulem}
\usepackage[squaren, Gray, cdot]{SIunits}

\def\jpm{j_{\pm}}
\def\kpm{\kappa_{\pm}}

\def\dotp{\dot{p}}
\def\tauT{\tau_{\rm T}}
\def\epsB{\varepsilon_{B}}

\def\rcoll{R_{\rm c}}

\def\Lrad{L_{\rm rad}}

\def\sigmaT{\sigma_{\rm T}}

\def\tauT{\tau_{\rm T}}

\def\ginj{\gamma_{\rm inj}}

\def\Urad{U_{\rm rad}}

\def\epsB{\varepsilon_{\rm B}}

\def\Lth{L_{\rm th}}
\def\Lnth{L_{\rm nth}}

\def\kth{k_{\rm th}}
\def\knth{k_{\rm nth}}

\def\Lrad{L_{\rm rad}}

\def\Inu{I_{\nu}}
\def\jnu{j_{\nu}}

\def\knu{\kappa_{\nu}}
\def\d{\partial}

\def\Zpm{Z_{\pm}}

\def\npm{n_{\pm}}

\def\rW{R_{\rm W}}

\def\enthinit{\varepsilon_{0,{\rm nth}}}
\def\ethinit{\varepsilon_{0,{\rm th}}}

\def\Epk{E_{\rm pk}}
\def\Rph{R_\star}

\def\Eq{Equation}

\def\sT{\sigma_{\rm T}}
\def\beq{\begin{equation}}
\def\eeq{\end{equation}}
\def\Ep{E_{\rm pk}}

\def\RW{R_{\rm W}}
\def\Rph{R_\star}

\def\dN{\dot{N}_\gamma}
 \def\gampeak{\gamma_{\rm pk}}
\def\Eav{\bar{E}}
\def\Sect{Section}
\def\Sects{Sections}
\def\Eq{Equation}
\def\Eqs{Equations}

\def\Lrad{L_\gamma}
\def\ep{\epsilon}

\def\TC{T_{\rm C}}

\def\thb{\theta_c}

\def\Teff{T_{\rm eff}}

\def\Ein{E_0}
\def\Emin{E_{\min}}

\def\lph{l_{\rm ph}}
\def\lCoul{l_{\rm Coul}}
\def\lln{\ell_n}
\def\rhof{\tilde{\rho}}
\def\wu{w_u}
\def\sigu{\sigma_u}
\def\rhofu{\tilde{\rho}_u}
\def\eps{\epsilon}

\newbox\grsign \setbox\grsign=\hbox{$>$} \newdimen\grdimen \grdimen=\ht\grsign
\newbox\simlessbox \newbox\simgreatbox \newbox\simpropbox
\setbox\simgreatbox=\hbox{\raise.5ex\hbox{$>$}\llap
     {\lower.5ex\hbox{$\sim$}}}\ht1=\grdimen\dp1=0pt
\setbox\simlessbox=\hbox{\raise.5ex\hbox{$<$}\llap
     {\lower.5ex\hbox{$\sim$}}}\ht2=\grdimen\dp2=0pt
\setbox\simpropbox=\hbox{\raise.5ex\hbox{$\propto$}\llap
     {\lower.5ex\hbox{$\sim$}}}\ht2=\grdimen\dp2=0pt
\def\simgt{\mathrel{\copy\simgreatbox}}
\def\simlt{\mathrel{\copy\simlessbox}}

\begin{document}

\title{Photospheric Emission of Gamma-Ray Bursts
}


\author{A. M. Beloborodov        \and
        P. M\'esz\'aros 
}


\institute{A. M. Beloborodov \at
              Physics Department and Columbia Astrophysics Laboratory, Columbia University, 538 West 120th Street, New York, NY 10027, USA  \\
              \email{amb@phys.columbia.edu}           
           \and
           P. M\'esz\'aros \at
           Center for Particle and Gravitational Astrophysics,
           Dept. of Astronomy \& Astrophysics and Dept. of Physics,
          Pennsylvania State University, University Park, PA 16802, USA \\
          \email{nnp@psu.edu}
}

\date{Received: date / Accepted: date}

\maketitle

\begin{abstract}
We review the physics of GRB production by  relativistic jets
that start highly opaque near the central source and then expand to transparency.
We discuss dissipative and radiative processes in the jet and how radiative transfer 
shapes the observed nonthermal spectrum released at the photosphere. 
A comparison of recent detailed models with observations gives estimates for 
important parameters of GRB jets, such as the Lorentz factor and magnetization.
We also discuss predictions for GRB polarization and neutrino emission.
\keywords{Gamma-Ray Bursts \and Radiative processes}
\end{abstract}

\section{Introduction}
\label{intro}

The emission mechanism of gamma-ray bursts (GRBs) has been a puzzle since
their discovery half a century ago. Following the discovery of GRB afterglows
in the end of 1990s, it has been established that GRB emission 
arises in ultra-relativistic jets from extremely powerful cosmic explosions at cosmological 
distances. The existing emission
models are mainly guided by the observed spectrum of the jet radiation
and its fast variability.

The main phase of a GRB, usually called ``prompt'' emission, typically lasts seconds
or minutes, with huge luminosities, sometimes reaching $L_\gamma\sim 10^{54}$~erg~s$^{-1}$
in isotropic equivalent (before correcting for the beaming of radiation from a collimated 
relativistic jet). The prompt phase should be distinguished from the GRB ``afterglow'' ---
the long-lasting and weaker emission with a broad spectrum extending from gamma-rays 
to radio waves. 

The spectrum of prompt emission has a well defined, sharp peak at an energy $\Ep$ 
that varies around 1~MeV, after correcting by $1+z$ for the cosmological redshift
\citep{Kaneko+06batsespec,Goldstein+12fermispcat}.  For most bursts the 
spectrum around the peak can be approximately described by a simple 
Band function \citep{Band+09} ---  two power laws that are smoothly connected 
at $\Ep$. Bursts of higher luminosity are observed to have higher $\Ep$. 
An approximate correlation $\Ep\approx 0.3\,L_{\gamma,52}^{1/2}$~MeV was reported
(e.g.  \cite{Wei+03-EpkLum,Yonetoku+04-EpkLum,Ghirlanda+12-Gamma-eiso}),
where $L_{\gamma,52}$ is the burst luminosity (isotropic equivalent) in units of 
$10^{52}$~erg~s$^{-1}$.
 
The short durations of $\sim 0.1-10$~s and fast variability of 
GRB emission imply that it is produced by outflows or jets with Lorentz factors 
{$\Gamma\simgt 10^2$. The Doppler effect compresses the typical observed 
timescale of radiation released at a radius $r$, from $\sim r/c$ to $\sim r/c\Gamma^2$.
In addition, a high $\Gamma$ allows the radiation to decouple from the jet 
at smaller radii; typically, the jet becomes transparent to MeV radiation at radii
somewhere between $10^{12}$ and $10^{14}$~cm.
The characteristic scattering optical depth seen by photons 
propagating from a radius $r$ to infinity is given by
\beq
\label{eq:tauT}
   \tauT=\frac{Z_\pm \dot{M}_p\sT}{4\pi r\Gamma^2m_pc},
\eeq
where $\dot{M}_p$ is the isotropic equivalent of the proton mass outflow rate,
and $Z_\pm$ is the number of $e^\pm$ per proton.
The optically thin zone of the jet roughly corresponds to $\tauT<1$, and the 
sub-photospheric region is where $\tauT>1$. Strong energy dissipation must occur in 
the jet to generate the observed non-thermal  spectra. This dissipation can occur 
before and after the jet becomes transparent to radiation.

\subsection{Optically thin synchrotron emission?}

A simple phenomenological GRB model posits that we observe Doppler-shifted 
synchrotron radiation, similar to blazar jets. The model assumes that a 
nonthermal electron population is injected in the jet by some dissipative process in 
an optically thin outflow. It gives a radiation spectrum that peaks at the photon energy
\beq
\label{eq:synch}
  \Ep\approx E_{\rm syn}= 0.4 \Gamma\,\gampeak^2\,\hbar\,\frac{eB}{m_ec}.
\eeq
Here $B$ is the magnetic field measured in the rest frame of the jet (``fluid frame''),
$\gampeak$ is the Lorentz factor at which the injected electron distribution 
peaks (also measured in the fluid frame), and $\Gamma$ is the bulk Lorentz 
factor of the jet itself. If the electron distribution at $\gamma_e>\gampeak$ is a 
power-law $dN_e/d\gamma_e\propto \gamma_e^{-p}$ then 
the synchrotron spectrum has a high-energy power-law tail,
$dN_{\rm ph}/dE\propto E^{-p/2-1}$ at $E>\Ep$ .

One possibility for the injection of high-energy electrons is associated with 
internal shocks \citep{Rees+94is,Kobayashi+97variab,Daigne+98is}.
A mildly relativistic electron-ion shock produces an electron distribution with
$\gampeak\sim \ep_e(m_p/m_e)$, where $\ep_e$ can be a significant fraction
of unity. This gives
\beq
\label{eq:synch1}
    \Ep\approx 1\,r_{12}^{-1}\ep_B^{1/2} L_{52}^{1/2}\left(\frac{\ep_e}{0.3}\right)^2
            {\rm ~MeV}, 
\eeq
where $r_{12}$ is radius in units in $10^{12}$~cm, $L_{52}$ is the isotropic 
equivalent of the jet power in units of $10^{52}$~erg~s$^{-1}$, and $\ep_B$
is the fraction of jet energy that is carried by the magnetic field. If the shock 
radius happens to be $r\sim 10^{13}\,\ep_B^{1/2}\ep_e^2$~cm then 
$\Ep$ would be consistent with observations. 
A Band-type spectrum is emitted if the electron energy distribution extends as 
a power law at $\gamma_e>\gampeak$ and cuts off at $\gamma_e<\gampeak$,
with a negligible Maxwellian component (e.g. \cite{Burgess+11issy}).
This is a problematic requirement.
In observed shocks, e.g. in the solar system or in supernovae, as well as in computer-simulated 
shocks, most of the dissipated energy is given to Maxwellian particles, with 
only a small fraction of particles populating the high-energy nonthermal tail.

Strong magnetization $\epsB\simgt 0.1$ offers an alternative mechanism
for particle acceleration: the dissipation of magnetic energy, in particular through magnetic 
reconnection  \citep{Drenkhahn+02}. Recent dedicated numerical studies 
of relativistic reconnection show that it gives a broad flat electron distribution, 
with no cutoff at low energies \citep{Kagan+15recon}. Synchrotron emission from this
electron distribution does not resemble the pronounced spectral peak observed in GRBs.

The synchrotron model may be viewed more broadly as a phenomenological
model which does not specify how the desired electron distribution is created.
In this form the model still faces a few problems: 
\\
(1) Thousands of GRBs have been observed, and most of them have $\Ep$
near 1~MeV (Goldstein et al. 2012). Few bursts have $\Ep$ above 10~MeV 
and no bursts are known with $\Ep>20$~MeV. This is at odds with 
the naive model prediction $\Ep\propto \Gamma\gampeak^2 B$, which  
should give a broad distribution of $\Ep$ due to variations in 
the local $B$, $\gampeak$, and $\Gamma$.
\\
(2) High-energy electrons quickly cool to $\gamma_e<\gampeak$ (which makes
the process radiatively efficient) and should emit radiation at $E<\Ep$ with 
a photon index $\alpha=-3/2$. A typical low-energy index observed
in GRBs is $\alpha\sim -1$, and many bursts have even harder slopes $\alpha>0$
(Kaneko et al. 2006).
Such hard slopes are in conflict with the synchrotron model.
\\
(3) The synchrotron spectrum should have a rather broad peak, even when the source is 
assumed to have a uniform magnetic field $B$ and the power-law electron distribution 
is assumed to cut off at $\gamma_e<\gampeak$.
Attempts to reproduce GRB spectra with such 
idealized models gave acceptable fits for a small number of bursts 
\citep{Burgess+11issy,Burgess+14therm,Preece+14-130427}
and were found incompatible with observed spectra in the dominant majority of  GRBs 
\citep{Axelsson+15spwidth,Yu+15spwidth}.

\begin{figure*}
\hspace*{0.5cm}
  \includegraphics[width=0.83\textwidth]{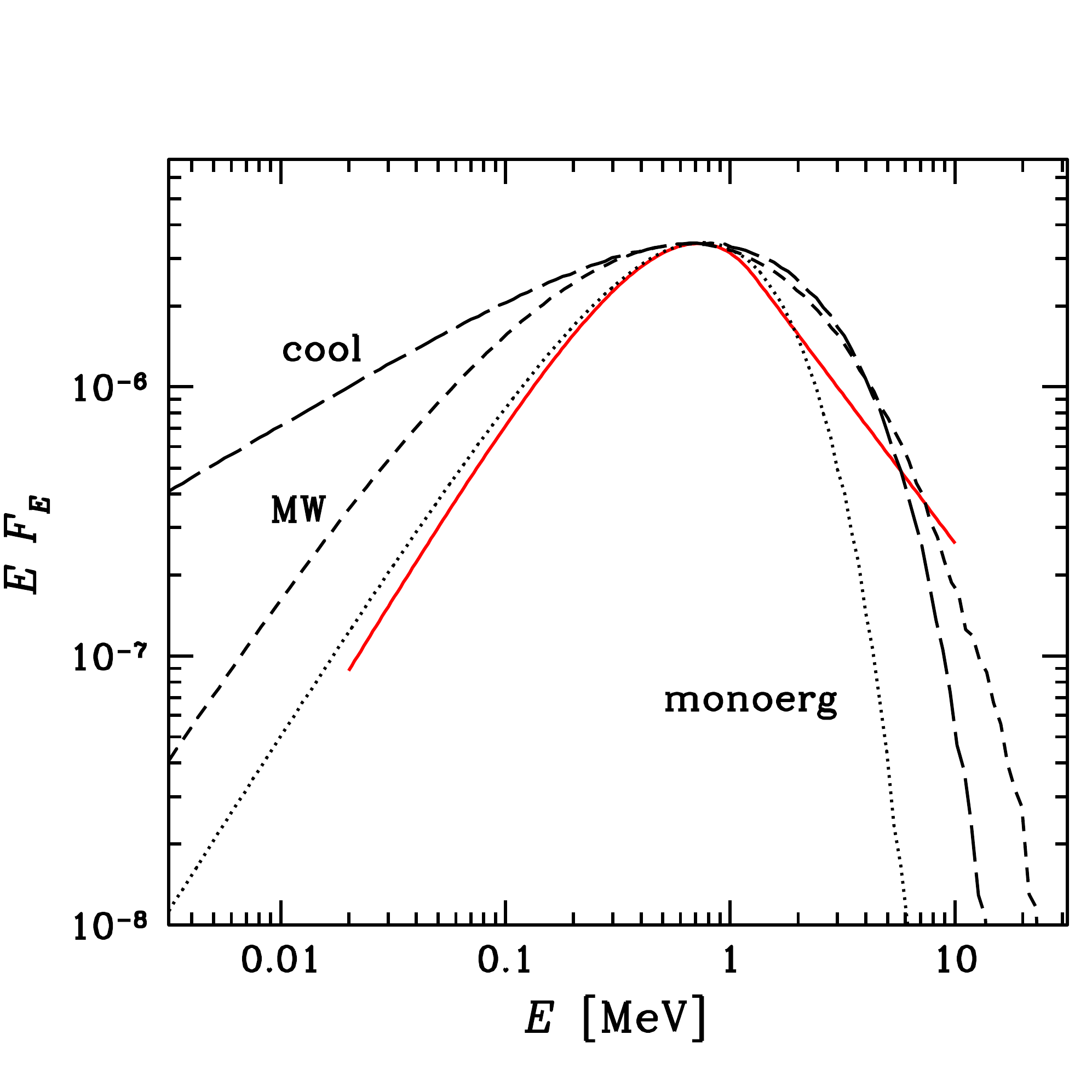}
\caption{Synchrotron spectra from an optically thin spherical shell.
Three models are shown with different electron distributions: 
mono-energetic (dotted line), Maxwellian 
(short-dashed line), and fast-cooling Maxwellian
(long-dashed line). For comparison the Band fit of GRB 990123 is shown by the red curve.
(From \cite{Vurm+16grbphot}.)
}
\label{fig:synch}       
\end{figure*}

The peak sharpness problem is illustrated in Figure \ref{fig:synch} where 
a synchrotron spectrum is compared with the Band fit of
a bright burst with a typical spectrum,  GRB~990123.
In the realistic fast-cooling regime, the minimum width of the synchrotron peak 
at half maximum exceeds $1.5$ orders in photon energy, even with a single value of $B$
in the emitting source and the narrow injected electron distribution (single-temperature
Maxwellian). It is significantly broader than the Band fit to the {\it time-average} of the 
GRB spectrum \citep{Briggs+99-990123}.
Uncertainties in the observed spectrum due to the detector response and limited
photon statistics allow some room for stretching the measured peak width;
however it can hardly be made consistent with the synchrotron model, 
especially when the inevitable broadening due to variable magnetic fields, 
electron injection and the Doppler factor are taken into account.
Note also that synchrotron spectra with sharp peaks consistent with
a Band function are not observed in any other astrophysical objects. 
One example is provided by blazars --- their synchrotron spectra have the 
half-maximum width of several orders of magnitude, broader than in GRBs 
(e.g. \cite{Ghisellini+06blazgrb}).

\subsection{Photospheric emission}

The sharp MeV spectral peak provides evidence for 
thermalization of radiation at early, opaque stages of the GRB explosion
(\cite{Paczynski86,Goodman86grb,Thompson94,Meszaros+00phot}.
The inheritance of the spectral peak from an initial thermalization stage is also 
supported by the observed distribution of $\Epk$, which cuts off above $\sim 10$~MeV
in agreement with a theoretical maximum \citep{Beloborodov13peak}.

Accepting that the MeV peak forms inside an opaque jet
leads to so-called ``photospheric'' emission models: the GRB radiation is released 
where the jet becomes transparent to scattering.
Several versions of the photospheric model have been discussed over the years
\citep{Thompson94,Eichler+00thermal,Rees+05photdis,Giannios+07photspec,
Beloborodov10pn,Levinson12radshock,Thompson+14phot}.
All share a key feature: the jet is {\it dissipative}, i.e. significantly heated as it 
propagates away from the central engine. 
This heating modifies the emitted photospheric radiation from simple blackbody 
emission. The resulting spectrum was investigated using dedicated numerical
simulations and found to have 
a nonthermal shape that closely resembles the phenomenological Band function
(\cite{Peer+06phot,Giannios08phot,Beloborodov10pn,Vurm+11phot,
Gill+14pairphot,Vurm+16grbphot}.
It was proposed that the dissipative photosphere provides the best
description for the observed spectra (e.g. \citealt{Ryde+11phot}).

Previous observational search for photospheric emission mainly looked for 
a Planck component or its modifications, see \citet{Peer+16} for a recent review. 
In contrast, the detailed theoretical calculations suggest that photospheric emission 
has a broad Band-like spectrum that dominates the prompt GRB rather than adds 
a blackbody component to it.

\section{Sub-photospheric dissipation}

Internal bulk motions and magnetic fields provide energy reservoirs available for dissipation
in GRB jets. Three mechanisms can make photospheric emission nonthermal: 
\begin{enumerate}
\item
Dissipation of magnetic energy 
\item
Collisions of drifting neutrons
\item 
Internal shocks
\end{enumerate}
The details of magnetic dissipation are complicated and still poorly understood. 
The other two dissipation mechanisms can be examined from first principles and 
give predictions that can be compared with observations; we will discuss them is 
some detail below.

A neutron component is expected in GRB explosions
\citep{Derishev+99,Bahcall+00pn,Meszaros+00gevnu,Beloborodov03neutron,2006MNRAS.369.1797R}.
Free neutrons enter the jet from the neutronized central engine and are also produced 
by spallation reactions in the jet itself.
They begin to drift with respect to the proton flow in the sub-photospheric region
and collide with the protons. These nuclear collisions create pions, whose decay leads to 
the production of $e^\pm$ pairs with Lorentz factors 
$\gamma_e\sim m_\pi/m_e\sim 300$ in the jet rest frame and to
 the emission of neutrinos with energy $\sim \Gamma m_\pi c^2$ in the observer frame.
These multi-GeV neutrinos are detectable with IceCube 
\citep{Bartos+13pn,Murase+13subphotnu}
and can provide an important test for GRB models.
The photospheric spectrum generated by neutron collisions has been calculated by 
\cite{Beloborodov10pn} and \cite{Vurm+11phot}, and found to be consistent with GRB spectra.

Internal shocks are generally expected to develop in GRB jets, as the 
jets are highly variable. The central engine of the explosion is likely unsteady,
and additional variability is induced as the jet breaks out of the progenitor star 
\citep{Lazzati+09phot,Lazzati+13dur,Ito+15photjet}.
This breakout is accompanied by multiple internal and recollimation shocks.
Shock heating is expected to occur in an extended range of radii and in an extended 
range of timescales, which is consistent with the broad power spectrum of variability 
observed in GRB light curves \citep{Beloborodov+00pdsgrb,Morsony+10variab}.

As long as the flow is opaque and radiation dominates its energy density, 
photon diffusion plays a leading role in shaping the shock front.
Its thickness is then comparable to the photon mean free path. 
The radiation-mediated shock (RMS) is not capable of 
electron acceleration by the standard Fermi mechanism, since the electron 
radiates its energy much faster than it can cross the shock. Photons can
experience significant energy gains by crossing the shock back and forth multiple times
\citep{Blandford+81radshock2,Riffert88radshock,Levinson12radshock}.
This ``bulk Comptonization'' upscatters photons up to the MeV band (in the fluid frame).
Upscattering to higher energies is hindered by the energy loss due to electron recoil 
in scattering. Photon upscattering beyond $\sim 1$~MeV is also stopped by 
the absorption reaction $\gamma+\gamma\rightarrow e^-+e^+$,
and the produced mildly relativistic $e^\pm$ pairs quickly cool down due to 
Coulomb and inverse Compton (IC) losses.

\begin{figure*}
\includegraphics[width=0.9\textwidth]{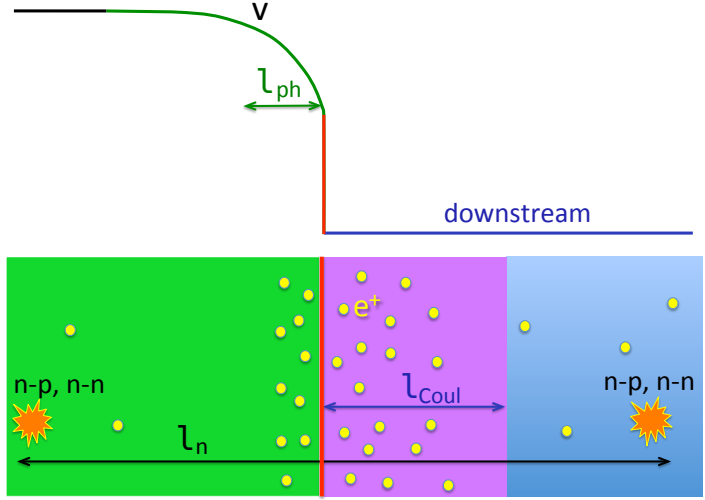}
\caption{Structure of a relativistic internal shock in a hot, opaque, magnetized outflow.
The green part of the velocity profile is shaped by radiation pressure on a scale
comparable to the photon mean free path to scattering, $\lph$. The vertical red part is 
the collisionless subshock, which heats ions to a mildly relativistic temperature and 
electrons to an ultra-relativistic temperature. 
The ultra-relativistic electrons are quickly cooled, producing synchrotron and 
inverse Compton (IC) photons, which create $e^\pm$ pairs ahead and behind the shock.
The ion cooling occurs on a longer scale $\lCoul$, which exceeds the RMS 
thickness if the subshock is relativistic. 
In the presence of a free neutron component in the flow, the shock wave is partially 
shaped by neutron migration between the upstream and downstream.
The neutron mean free path $\lln$ is the longest scale in the shock structure. 
The migrating neutrons are stopped by nuclear collisions.
Inelastic nuclear collisions result in injection of $e^\pm$ with 
Lorentz factors $\gamma_e\sim m_\pi/m_e\sim 300$, 
distributed over a broad region $\sim \ell_n$.
The nuclear collisions also emit neutrinos; they escape the outflow with energies 
$\sim \Gamma m_\pi c^2$ where $\Gamma$ is the outflow Lorentz factor.
(From \cite{Beloborodov16phot}.)
}
\label{fig:scheme}       
\end{figure*}

All this would suggest that sub-photospheric shocks are inefficient in producing 
particles with energies $E\gg m_ec^2$ in the fluid frame. However, 
a more realistic picture of sub-photospheric shocks differs from the simple RMS,
in particular in outflows that carry magnetic fields and free neutrons
in addition to the plasma and radiation.
The shock wave is capable of generating ultra-relativistic electrons in two ways:
\begin{itemize}
\item
A strong collisionless subshock forms in the RMS. This is 
inevitable (even deep below the photosphere) if the flow is sufficiently magnetized
\citep{Beloborodov16phot}.
A mildly relativistic collisionless subshock heats the electrons to 
an ultra-relativistic temperature $T_e$, as they receive a significant fraction of 
the ion kinetic energy (e.g. \citealt{Sironi+11magshockei}).
One can show that the inverse Compton (IC) emission from the heated electrons
breeds $e^\pm$ pairs in the upstream, 
and this process regulates the subshock temperature to $kT_e\sim 10 m_ec^2$.
\item
Inelastic nuclear collisions inject $e^\pm$ pairs with Lorentz factors 
$\sim m_\pi/m_e\sim 300$ in the fluid frame. This mechanism 
becomes particularly efficient if the outflow carries free neutrons, as they can migrate 
across the RMS, making the shock wave partially mediated by neutrons (Figure~\ref{fig:scheme}).
\end{itemize}

The structure of sub-photospheric shocks can be studied using 
detailed numerical simulations. One method is to
search for a steady-state propagating solution of the kinetic equations
(e.g. \cite{2008PhRvL.100m1101L,Budnik+10radshock}), which may be found by iterations.
Alternatively, one can use direct time-dependent simulations of shock formation.
The simulation can be set up to follow the evolution of 
a compressive wave in the outflow, which leads to formation of a pair of shocks and 
their subsequent quasi-steady propagation.
Then the shock structure is obtained from first principles by calculating the 
time-dependent radiative transfer through the moving plasma.
A detailed first-principle simulation involves tracking of a large
number of individual photons and their interaction with the moving plasma (using 
Monte-Carlo techniques). 

The shock structure obtained with this method
is shown in Figure~\ref{fig:shock}, for two cases --- with and without magnetic fields.
The simulations demonstrate the formation of a strong collisionless subshock in the 
magnetized RMS. In GRB jets, a moderate magnetization $\sim 0.1$ is sufficient to 
generate a strong subshock. 

\begin{figure*}
  \includegraphics[width=0.5\textwidth]{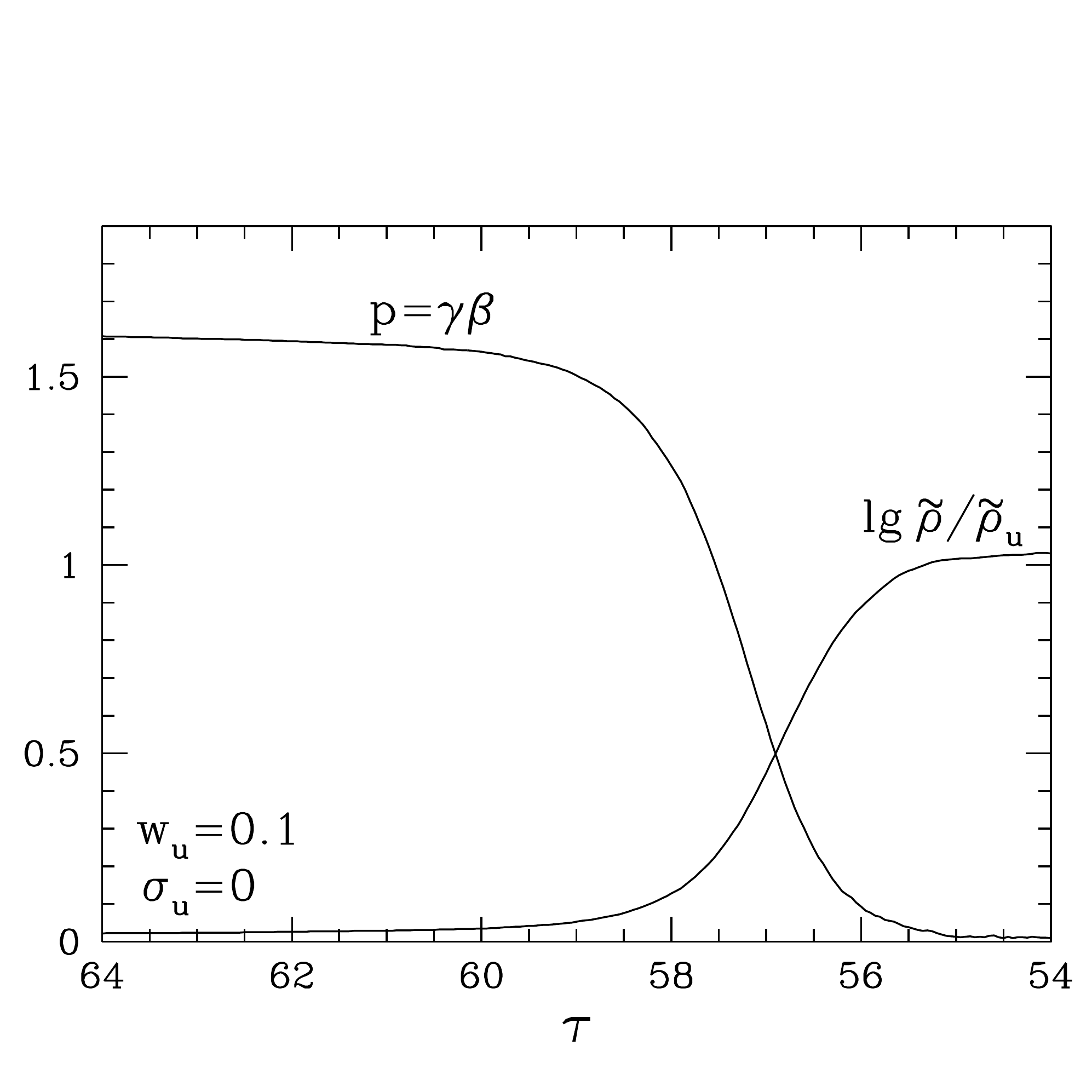}
   \includegraphics[width=0.5\textwidth]{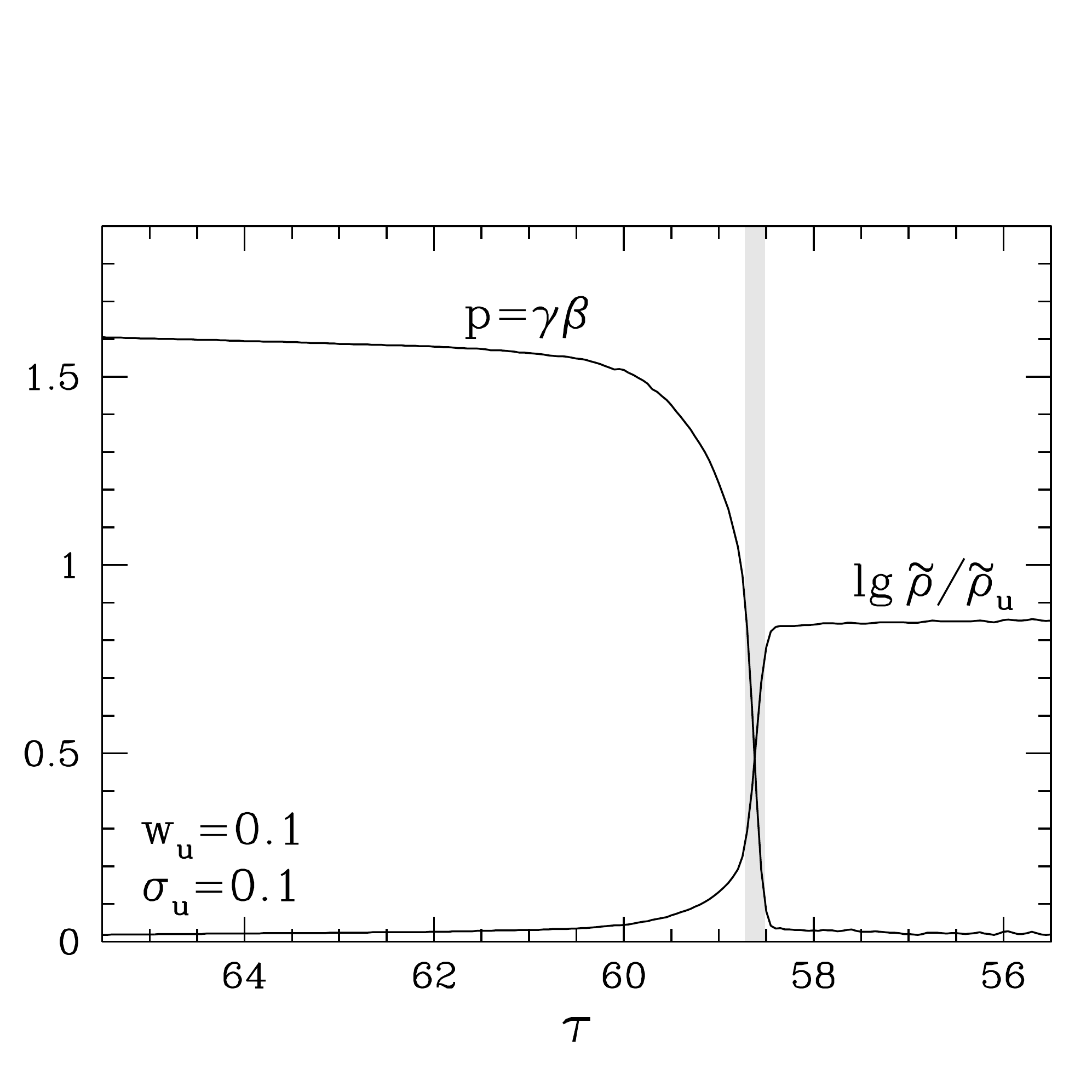}
\caption{Snapshot of a shock propagating in the flow with upstream radiation
enthalpy $\wu=0.1$ and upstream magnetization $\sigu=0$ (left) and 0.1 (right).  
The parameters $w$ and $\sigma$ are defined in the fluid frame as 
$w=(4/3)\Urad/\rhof c^2$ and $\sigma=B^2/4\pi\rhof c^2=2\epsB$, where 
$\Urad$ is the radiation energy density and $\rhof$ is the proper mass density of the fluid.
The shock is propagating to the left and the Thomson optical depth $\tau$ is 
measured from the site of shock formation (caustic of the initial supersonic wave).
The solid curves show the profiles of momentum $p=\gamma\beta$ and 
proper density $\rhof$ (normalized to the upstream proper density $\rhofu$).
A strong subshock has formed in the magnetized case; it is highlighted by the grey strip.
The electron temperature has a strong and narrow peak at the subshock.
The subshock is resolved (not a discontinuous jump) due to a finite viscosity employed in 
the simulation of the plasma dynamics. 
The radiation is everywhere simulated directly as a large collection of individual photons 
whose propagation and scattering is followed using the Monte-Carlo technique. 
(From \cite{Beloborodov16phot}.)
}
\label{fig:shock}       
\end{figure*}

Electron-positron pair creation was expected in dissipative GRB jets since the 
early works in the 1990s (e.g. \cite{Rees+94is,Thompson94,Meszaros+00phot}).
The efficiency of this process is described by the pair loading factor $Z_\pm=n_\pm/n_p$ 
where $n_p$ is the proton density and $n_\pm$ is the $e^\pm$ density.
One can show that $Z_\pm$ reaches a peak of $\sim 10^2$ inside the shock front 
and decreases behind it due to pair annihilation \citep{Beloborodov16phot}. 
The pairs are produced in collisions between MeV photons, which 
are generated by two mechanisms:
(1) An RMS without a strong collisionless subshock
produces MeV photons due to bulk Comptonization by a relatively cold
flow (the flow is everywhere in local Compton equilibrium with radiation at 
$kT\ll m_ec^2$). 
(2) In the presence of a collisionless jump, MeV photons are produced by IC cooling 
of $e^\pm$ heated to the subshock temperature $kT_e\sim 10 m_ec^2$.

The shock ``breakout''  at the photosphere\footnote{The photospheric shock breakout 
     should not be confused with the jet breakout from the progenitor star. The jet 
     breakout occurs at smaller radii and huge optical depths.}
occurs through the growth of the collisionless jump in the RMS until radiation 
completely decouples from the plasma. Eventually the entire velocity jump becomes 
mediated by collective plasma processes, regardless of magnetization.
This somewhat resembles the shock breakout in non-relativistic supernova explosions 
\citep{Waxman+01nusn,Giacinti+15nusn}.
However, the photospheric shock breakout in GRBs has a special feature:
$e^\pm$ pair creation. The shock sustains a local optical depth 
$\tauT\simgt 1$ even after the background electron-ion plasma becomes transparent.
Effectively, the shock carries the photosphere with it until it expands by an additional
factor $\sim 30-100$, continually producing photospheric emission \citep{Beloborodov16phot}.
This emission should be observed as a prominent pulse of nonthermal radiation in 
the GRB light curve. In the observer frame, the emission from the $e^\pm$-dressed 
shocks extends up to the GeV band,
where $\gamma$-$\gamma$ absorption shapes a break in the spectrum. 

The high-energy photospheric pulses contribute to the prompt GRB emission. 
This emission is observed to overlap (in arrival time) with GeV afterglow, which is 
associated with the external blast wave from the explosion \citep{Ackermann+13-110731}. 
The early peak of GeV afterglow is explained by the $e^\pm$-loading of the blast 
wave \citep{Beloborodov+14pairgev,Hascoet+15grblf}, and its overlap with the 
high-energy photospheric pulses can explain the observed 
variable component of GeV emission at early times.

Internal shocks are the main heating mechanism for jets that are not dominated by 
magnetic fields. This is a likely situation in GRBs, as 
comparison of detailed models of photospheric radiation with observed GRB spectra 
suggests a moderate magnetization in the sub-photospheric region, 
$\sigma\sim 10^{-2}-10^{-1}$ (see \Sect~\ref{spectrum} below). This does not  
exclude a stronger magnetization closer to the central engine, allowing for magnetic
dissipation that reduces $\sigma$ as the jet expands. 
In this scenario, the early heating at very large optical depths may be
dominated by magnetic dissipation. Constraints on the jet magnetization 
at small radii were recently estimated using the jet breakout time by 
\cite{Bromberg+15jetcomp}; they find that the jet is not dominated by magnetic fields.

Neutron migration across the shock front leads to nuclear collisions which generate
additional $e^\pm$ pairs and produce neutrino emission.
The typical observed energy of neutrinos produced by this mechanism is 
$\sim\Gamma m_\pi c^2\simgt 10$~GeV.
Sub-photospheric internal shocks were also proposed to emit ultra-high-energy neutrinos 
\citep{Meszaros+01choked,Razzaque+03nutomo}.
This proposal assumes efficient ion acceleration by the Fermi diffusive mechanism,
which does not operate in an RMS and requires a collisionless (sub)shock.
Diffusive acceleration is, however, suppressed in shocks with a 
transverse magnetic field, which advects the particles downstream before they 
have a chance to cross the shock many times \citep{Sironi+11magshockei}.
Obliqueness of the magnetic field could help the ion acceleration.
Another possible way to accelerate ions is the converter mechanism proposed by 
\cite{Derishev+03converter}. Numerical simulations of \cite{Kashiyama+13pnconv} 
suggest that the converter mechanism becomes efficient for ultra-relativistic shocks, 
with amplitude $\gamma_0\simgt 4$. This mechanism is, however, quite slow as it 
relies on neutron-to-proton and proton-to-neutron conversion in (inelastic) nuclear collisions, which occur with a large free path $l_n\sim (n_p\sigma_n)^{-1}$ due to the modest 
nuclear cross section $\sigma_n\approx \sT/20$.


\section{Peak position of the photospheric spectrum $\Ep$}
\label{peak}

Photons deep below the photosphere keep interacting with the plasma, 
and their spectrum forms a sharp Wien peak.
The simplest model of photospheric emission assumes a freely expanding 
radiation-dominated outflow with no baryon loading or magnetic fields 
\citep{Paczynski86,Goodman86grb}. In this case, dissipation weakly influences 
the radiation spectrum and it is not far from a Planck shape.
The peak energy of the Planck spectrum (defined as the peak of $dL/d\ln E$) 
is related to the average photon energy $\Eav$ by $\Ep\approx 1.45\Eav$.
In the ideal radiation-dominated outflow $\Eav$ remains constant and equal
to its value near the central engine $\Ein$.

This simple dissipationless model, however, fails to explain the observed spectra.
Although the Planck spectrum may appear in the time-resolved emission of some 
bursts (e.g. \cite{Ryde04,Ryde+11phot}), GRB spectra are typically nonthermal, with 
an extended high-energy tail.

\subsection{Effects of collimation and early dissipation}

The average photon energy at the explosion centre, $\Ein$, is a useful parameter of
more general jets, which may be dissipative, magnetized, and collimated. $\Ein$ may 
be expressed in terms of the jet power $L_0$ and an initial radius $r_0$ 
(comparable to the size of the central compact object) using the relation 
$\ep_0 L_0\approx 4\pi r_0^2 aT_0^4 c$,
\beq
\label{eq:E0}
   \Ein\approx 10\, \ep_0^{1/4}\,L_{0,52}^{1/4}\, r_{0,6}^{-1/2}~{\rm MeV}.
\eeq
Here $\ep_0$ is the initial thermal fraction of the power $L_0$
(radiation-dominated jets with sub-dominant magnetic energy have $\ep_0\approx 1$).
Collimation of the flow within a small angle $\thb$ implies a smaller true jet power 
$L_0$ compared with its apparent isotropic equivalent $L_\gamma$,
\beq
   L_0\approx\frac{\thb^2}{2}\, L_\gamma,
\eeq
and hence $\Ein$ is reduced as $\thb^{1/2}$.
Achromatic breaks in GRB afterglow light curves are often interpreted as 
evidence for jet collimation, with a typical opening angle of 5-10$^{\rm o}$.
Collimation helps explain the extremely high apparent luminosities $L_\gamma$,
which reach $10^{54}$~erg~s$^{-1}$ in some GRBs. 
This collimation can be the result of the pressure confinement 
of the jet by the progenitor star and the breakout cocoon, or by a 
dense wind from the outer regions of the central accretion disk.
Collimation may also be assisted by strong magnetic fields \citep{Komissarov+09maggrb}.

If collimation is not accompanied by significant dissipation, the expanding
jet can be described as an ideal relativistic flow 
confined by a wall that determines the cross section of the jet.
Dissipationless collimation conserves entropy, and hence does not change
the photon-to-baryon ratio $n_\gamma/n$. This implies 
conservation of the photon number carried by the jet and a constant energy per photon
$\Eav$. Collimation boosts the isotropic equivalent of the luminosity
$L_\gamma$ and the isotropic equivalent of the photon flux $\dN$ 
by the same factor $\sim\thb^{-2}$, and their ratio 
$\Eav=\Lrad/\dN$ remains unchanged from its value at $r_0$, $\Eav=\Ein$.

In contrast, dissipative collimation (e.g. involving re-collimation shocks) 
generates entropy and hence increases $n_\gamma/n$.
Then the total photon number carried by the jet is increased by a factor of $Q>1$ 
and hence $\Eav$ (jet energy per photon) is reduced as $Q^{-1}$,
giving a GRB with a reduced $\Ep$.

If there is a relation between $\thb$ and $Q$, it leads to a correlated variation 
of $L$ and $\Ep$ (with $\thb$ being the varying parameter). 
\cite{Thompson+07phot} considered the possibility that $\thb$ always tends 
to its maximum value $\thb\sim\Gamma^{-1}$ allowed by the condition of causal
contact across the jet. They pointed out 
that this gives $\Eav\propto \thb^{-1}$ and hence $\Ep\propto L^{1/2}$,
similar to the observed trend. This simple explanation of the 1/2 slope of the 
$\Ep$-$L_\gamma$ correlation is however problematic, as
it invokes huge variations in $\thb$ to cover the observed correlation range, 
not only from burst to burst but also within a single burst (e.g. Ghirlanda et al. 2011).
It also assumes that GRBs of various apparent luminosities $L\sim L_0\thb^{-2}$ have 
approximately the same true power $L_0\sim 10^{50}$~erg~s$^{-1}$, which 
implies a central temperature $kT_0\approx 1\,r_{0,6}^{-1/2}$~MeV. 
Then the brightest bursts would have the highest 
$\Ep\approx 3kT_0\approx 3 r_{0,6}^{-1/2}$~MeV, which 
falls short of the observed highest $\Ep\sim 15$~MeV. 

\begin{figure*}
  \includegraphics[width=0.7\textwidth]{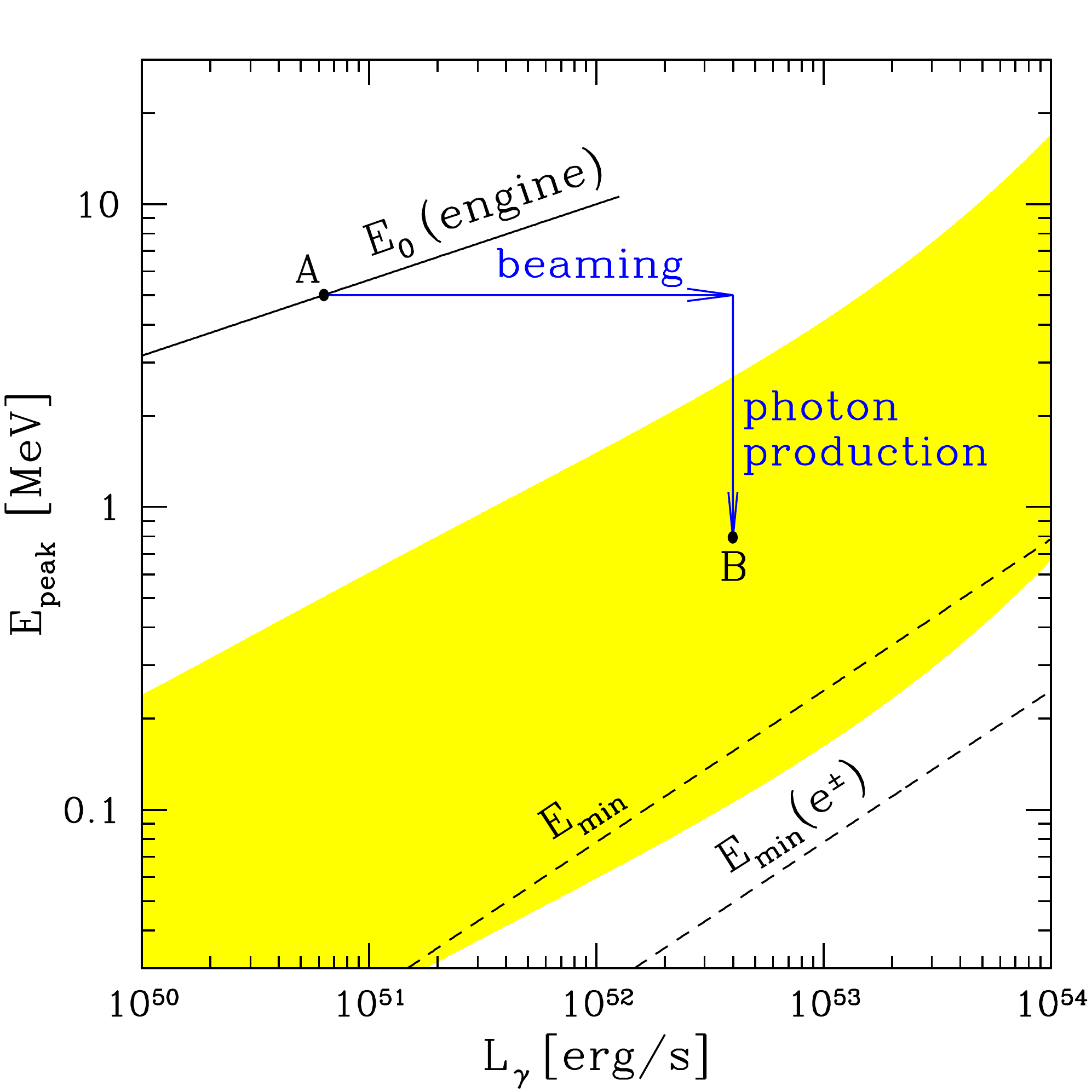}
\caption{$\Lrad$-$\Ep$ diagram. Point A is an example of the initial condition near the 
central engine; the jet starts with $\Ep$ close to $\Ein$ (\Eq~\ref{eq:E0}).
As the jet expands, its apparent luminosity $\Lrad$ is increased by beaming 
and $\Ep$ is reduced by photon production (dissipation offsets adiabatic cooling). 
Point B shows the resulting photospheric emission. The approximate region 
populated by observed GRBs 
is shown in yellow. The observed $\Ep$ should not violate the lower bound 
$\Emin$ that is set by the effective blackbody temperature of the photospheric 
radiation. The dashed lines show $\Emin$ 
with and without pair enrichment of the photosphere.
(From \cite{Beloborodov13peak}.) 
}
\label{fig:Ep}       
\end{figure*}

Although the dissipative collimation is not the only process regulating the observed 
$\Epk$ (as will be further discussed below), it plays an important role illustrated 
in Figure~\ref{fig:Ep}:
beaming/collimation increases the apparent luminosity $L_\gamma$ while
photon production due to dissipation reduces the observed $\Ep$.
Reasonable beaming factors $L/L_0\sim 10^2$ (suggested by the burst 
energetics and the afterglow data analysis) together with the expected photon 
production in a dissipative jet naturally explains the location of observed GRBs 
on the $\Lrad$-$\Ep$ diagram.

Evidence for dissipation at small radii and large optical depths is provided 
by the observed {\it photon number} emitted in GRBs.
In many GRBs, the blackbody
central engine is unable to provide the observed photon number, 
so additional photons must be produced in the expanding jet.

\subsection{Planck, Wien, and nonthermal zones below photosphere}

Observations also require that dissipation in many GRBs continues at least 
to the photospheric radius, so that the released spectrum has a nonthermal tail. 
The effect of dissipation on $\Ep$ differs in the following three 
sub-photospheric zones \citep{Beloborodov13peak}.

\begin{figure*}
  \includegraphics[width=1.0\textwidth]{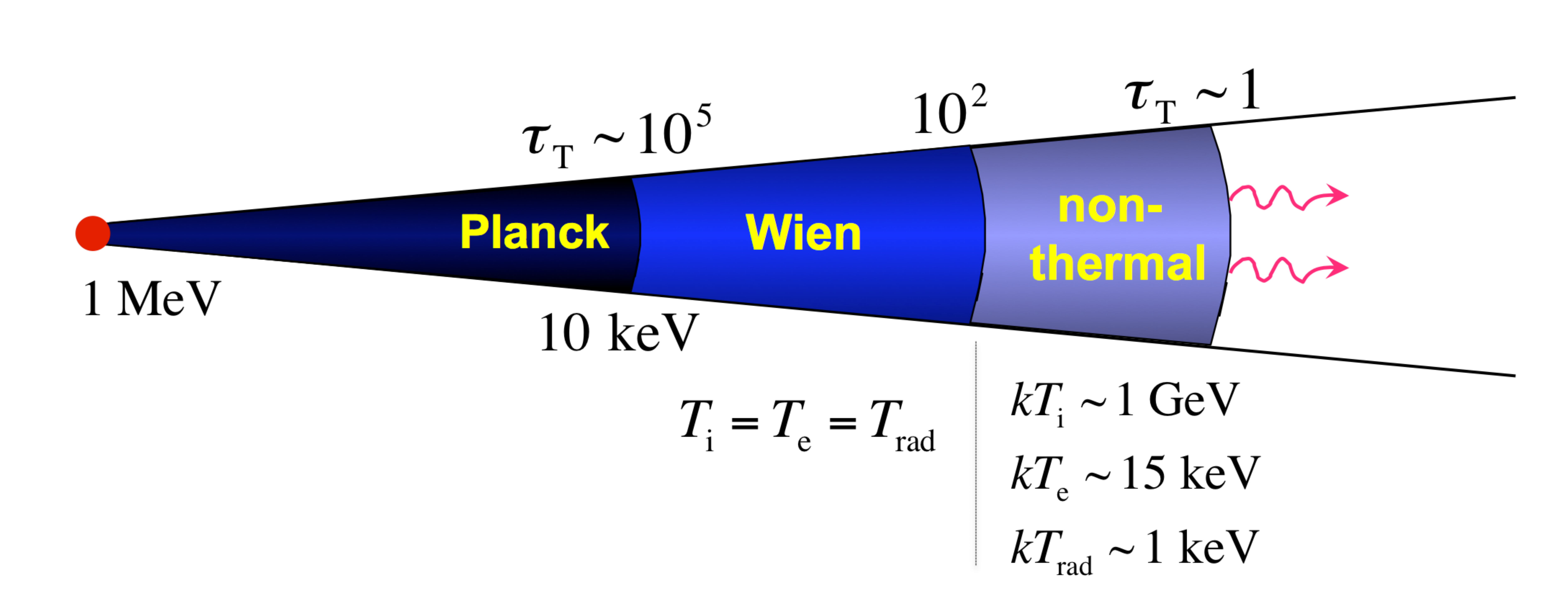}
\caption{
 Three zones in a GRB jet: 
 (1) Planck zone where a blackbody spectrum is enforced; this 
zone ends where the Thomson optical depth decreases to $\tauT\approx 10^5$. 
(2) Wien zone with Kompaneets parameter $y\gg 1$ where 
radiation has a Bose-Einstein spectrum, and (3) Comptonization 
(``nonthermal'') zone where the radiation spectrum develops the high-energy tail.
In the Planck and Wien zones the ion and electron temperatures are equal 
to the photon temperature, $T_i=T_e=T_{\rm rad}$. Outside the Wien zone 
$T_i\gg T_e\gg T_{\rm rad}$. The approximate values of the temperatures are 
indicated in the figure; these characteristic values weakly depend on how the 
jet is heated. The radiation spectrum is nonthermal in the Comptonization zone and 
here $T_{\rm rad}$ is defined as the Compton temperature, $T_{\rm rad}=T_{\rm C}$ 
(a measure of the mean photon energy).}
\label{fig:Wien}       
\end{figure*}

\begin{enumerate}

\item Planck zone ($r\simlt 10^{10}$ cm, $\tauT \simgt 10^5$):
the density of the jet is sufficiently high to maintain blackbody radiation in 
detailed equilibrium with the thermalized plasma. The thermal plasma efficiently 
produces photons through double Compton scattering.

\item Wien zone ($\tauT \simgt 10^2$): the dissipated heat is thermalized
into a Bose-Einstein photon distribution with a {\it finite chemical potential}.
Kinetic equilibrium between the photons and the plasma is maintained due to 
the large Kompaneets parameter of the thermal plasma, 
$y=4(kT/m_ec^2)\tauT\gg 1$.
The number of photons accumulated in the Wien peak saturates near the Wien radius
$\RW$ where $y$ drops to $\sim 1$
and Comptonization becomes unable to bring 
new generated photons to the spectral peak.

\item Unsaturated Comptonization zone ($1\simlt \tauT \simlt 10^2$): 
heating maintains a Compton parameter $y\sim 1$. The final non-thermal shape of the 
spectrum is produced in this region. 
Depending on the details of the dissipation mechanism, the spectrum can develop
a high-energy tail up to the GeV
band. In the presence of a soft photon source (synchrotron emission) unsaturated 
Comptonization also produces a low-energy photon index $\alpha\sim 1$.

\end{enumerate}
Dissipation in the Planck zone reduces $\Ep$. Dissipation in the Wien zone 
increases $\Ep$ if no new photons are produced;
strong synchrotron emission in the Wien zone can reduce $\Ep$.
In the non-thermal zone $\Ep$ remains practically unchanged.

A blackbody photosphere would have the temperature $T=\Teff$ defined by
\beq
\label{eq:Teff}
   \frac{4}{3}\,a\Teff^4\,\Gamma^2\,4\pi\Rph^2 \,c=L_\gamma.
\eeq
The corresponding $\Ep$ would be given by $\Ep\approx 4\Gamma\,k\Teff$. 
This estimate should be viewed as a robust lower limit on $\Ep$ rather than its true value.
One could also define the effective blackbody temperature at the Wien radius,
replacing $\Rph$ with $\RW$ in \Eq~(\ref{eq:Teff}) \citep{Giannios12peak}.
This would still give $\Ep$ lower than its true value.
A more realistic model should not make the blackbody assumptions at any radii. 
Instead, it should explicitly follow the production of photons (and their Comptonization) 
in the expanding jet. Such simulations have been performed recently; their results are
described below.
 
Radiative transfer simulations also give a detailed description of radiation decoupling
from the expanding jet. There is no unique sphere of last scattering, and the notion of 
photosphere is quite fuzzy. $\Rph$ is defined as the radius where $\tauT=1$ 
(see \Eq~\ref{eq:tauT}). This gives a characteristic decoupling radius, however the 
location of last scattering is random and surprisingly broadly distributed around $\Rph$ \citep{Peer_2008,Beloborodov11radtran}. 2/3 of photons diffusing from large optical 
depths are last scattered between $0.3\Rph$ and $3\Rph$, and 1/3 --- outside this 
interval. Another surprising transfer effect is the significant anisotropy of radiation 
(measured in the jet rest frame) developing at small radii $r\sim 0.1\Rph$ \citep{Beloborodov11radtran}.


\section{GRB spectrum from radiative transfer simulations}
\label{spectrum}

Radiation emerging at the photosphere
carries information about the entire expansion history of the jet, and
the observed spectrum may be used to reconstruct dissipative processes 
hidden in the opaque region behind the photosphere.
This can be achieved by solving a carefully formulated
radiative transfer problem, which
must be solved consistently with the flow dynamics and heating.
A global simulation starting at a sufficiently small radius will show 
how the jet radiation deviates from blackbody and becomes nonthermal,
shaping the observed GRB spectrum.

The effect of subphotospheric heating on GRB radiation has been studied with four 
different numerical codes 
\citep{Peer+06phot,Giannios08phot,Beloborodov10pn,Vurm+11phot},
consistently giving Band-type spectra. 
These calculations explored how the spectrum develops the high-energy tail 
at optical depths $\tau\simlt 30$ as a result of thermal and nonthermal Comptonization. 
Recently, \cite{Vurm+16grbphot} have performed global simulations of radiative 
transfer starting from very high optical depths $\tau>10^3$ and including photon 
production and absorption processes.

\subsection{Formulation of the transfer problem}

A convenient starting point of transfer simulations is the radius $R_c$ where the 
collimation process ends and the jet begins to expand conically.
The jet is still accelerating at this stage. 
The acceleration can be self-consistently determined using the dynamical equation 
\citep{Beloborodov11radtran}, 
\begin{align}
\frac{d\Gamma}{d r} =  \frac{\sigmaT \Zpm}{m_p c^3}\, F,
\label{eq:Gevol2}
\end{align}
where $F$ is the radiation flux measured in the plasma rest frame,
and $Z_\pm$ is the number of electrons and positrons per proton.
\Eq~(\ref{eq:Gevol2}) assumes that radiation pressure dominates the jet acceleration,
neglecting the effect of magnetic forces; this approximation may be reasonable even 
when the jet is strongly magnetized, as long as the jet is hot 
\citep{Vlahakis+03grbmhd1,Russo+13magjet1}. 

Radiation carried by the dissipative jet is influenced by electron heating.
Two types of heating should be distinguished: thermal and nonthermal:

\begin{itemize}

\item 
Thermal heating is the energy deposition into the Maxwellian $e^\pm$ plasma.
In particular, in a two-temperature plasma 
$T_i\gg T_e$ below the photosphere the electrons are efficiently heated by Coulomb 
collisions with the hot ions (Beloborodov 2010). 
Thermal heating tends to increase the $e^\pm$ temperature above the local Compton 
temperature of radiation $\TC$, and the decoupling $T_e>\TC$ becomes significant 
at $\tauT\simlt 10^2$.

\item
Nonthermal heating is the injection
of ultra-relativistic particles with a Lorentz factor $\ginj$, e.g. by inelastic nuclear 
collisions ($\ginj\sim m_\pi/m_e\sim 300$) or by other dissipation mechanisms discussed 
in \Sect~2. The energetic particles can generate an inverse Compton cascade and their 
energy is processed into secondary $e^\pm$ pairs of lower energies (Svensson 1987).
The value of $\ginj$ is only important for synchrotron emission, which dominates 
the observed spectrum at $E\ll\Ep$ (see below) and has little relevance to the main
MeV peak. Like thermal heating, the main parameter of nonthermal heating is its power.

\end{itemize}
To keep the number of parameters to a minimum, \cite{Vurm+16grbphot} 
employed a simple model of continuous internal dissipation,
approximating the thermal and nonthermal dissipation rates as power laws of radius, 
\beq
   \frac{d\Lth}{d\ln{r}} = \ethinit L \left( \frac{r}{\rcoll} \right)^{\kth}, \qquad
    \frac{d\Lnth}{d\ln{r}} = \enthinit L \left( \frac{r}{\rcoll} \right)^{\knth},
\eeq
where $L$ is the total energy flow rate in the jet. The power law may be a crude 
approximation to jet heating by multiple internal shocks or reconnection, 
however it allows one to study the global picture of the evolution of radiative 
processes with radius in the expanding jet. The formation of
photospheric radiation extends over several decades in radius, and
the global simulation allows one to see all stages of this process.

Radiation is described by its intensity $I_\nu(r,\theta,\nu)$ in the fluid frame,
where $\theta$ is the photon angle with respect to the radial direction.
The radiative transfer equation in ultra-relativistic outflows ($\Gamma>10$) 
is given by \citep{Beloborodov11radtran},  
\begin{align}
\nonumber
  \frac{1}{r^2 \Gamma}\frac{\partial}{\partial\ln{r}} \left[ (1+\mu) r^2 \Gamma \Inu \right] &= 
\frac{r}{\Gamma} (\jnu - \knu\Inu) + (1+\mu)(1 - g\mu) \frac{\d\Inu}{\d\ln\nu}	 \\
&  - \frac{\d}{\d\mu} \left[ (1-\mu^2)(1+\mu) \, g\Inu
\right],
\label{eq:RTE}
\end{align}
where $\mu = \cos\theta$, $\kappa_\nu$ is the opacity, $j_\nu$ is the emissivity, and 
\begin{align}
    g = 1- \frac{d\ln\Gamma}{d\ln{r}}.
\end{align}
\Eq~(\ref{eq:RTE}) describes a steady jet, i.e. a simplified model where 
variability and internal shocks are averaged out and replaced by dissipation terms.
At very large optical depths below the photosphere, $\tauT>100$, 
the radiation is approximately isotropic in the fluid frame, which simplifies the 
transfer problem --- it is reduced to advection of isotropic radiation by the expanding outflow. 
However, a new feature in this region is the importance of induced scattering.
Therefore, at $\tauT>100$ \Eq~(\ref{eq:RTE}) should be replaced by the Kompaneets equation (see \citealt{Vurm+13phot}; \citealt{Vurm+16grbphot}).
At optical depths $\tauT\ll 100$ the radiation becomes significantly anisotropic,
and the full transfer Equation~(\ref{eq:RTE}) should be used; the induced scattering 
rate is small at these radii and can be neglected.

The radiative transfer is coupled to the evolution of the electron/positron 
distribution in the expanding jet, which is described by the kinetic equation
\citep{Vurm+11phot},   
\begin{align}
&\frac{1}{r^2\Gamma} \frac{\d}{\d\ln{r}} \left[ r^2\Gamma \, \npm(p) \right] = 
\frac{r}{c\Gamma} (\jpm - c\kpm \npm(p)) \nonumber \\
& -\frac{\d}{\d p}
\left\{
\frac{r}{c\Gamma} \left[
\dotp\, \npm(p) - \frac{1}{2} \frac{\d}{\d \gamma}\left[ D \npm(p) \right]
\right]
- \frac{3-g}{3} \, p \,  \npm(p)
\right\}.
\label{eq:elkin}
\end{align}
Here $p=\sqrt{\gamma^2 - 1}$ is the electron/positron momentum in 
units of $m_ec$, and $\npm(p)$ is the $e^\pm$ momentum distribution.
The terms $\dotp$ and $D$ account for heating/cooling and diffusion in 
the momentum space due to radiative processes and Coulomb collisions.
The emission term $\jpm$ includes the injection of new pairs due to non-thermal 
dissipation and photon-photon absorption $\gamma+\gamma\rightarrow e^++e^-$, 
and $c\kappa_\pm n_\pm(p)$ is the pair annihilation rate.

Thermal radiative processes include Comptonization of photons by the thermal plasma,
induced Compton scattering at low frequencies, cyclotron emission/absorption, 
bremsstrahlung, and double Compton scattering. Nonthermal radiative processes 
result from the injection of relativistic leptons. This leads to the IC $e^\pm$ cascade
and synchrotron emission, which is an important source of photons.
All these processes have been followed in the expanding 
dissipative jet by \cite{Vurm+16grbphot} using the exact cross sections and rates
and solving numerically the transfer problem described by the coupled 
\Eqs~(\ref{eq:Gevol2}), (\ref{eq:RTE}), (\ref{eq:elkin}).

\subsection{Main features of sub-photospheric radiative transfer}

Figure \ref{fig:sp_evol} shows the evolution of the radiation spectrum within 
a dissipative jet as the jet expands from the highly opaque region $\tauT\sim 3\times 10^3$
to transparency. There are two main stages of the spectral evolution:

\begin{figure*}[t]
\vspace*{-1cm}
  \includegraphics[width=0.77\textwidth]{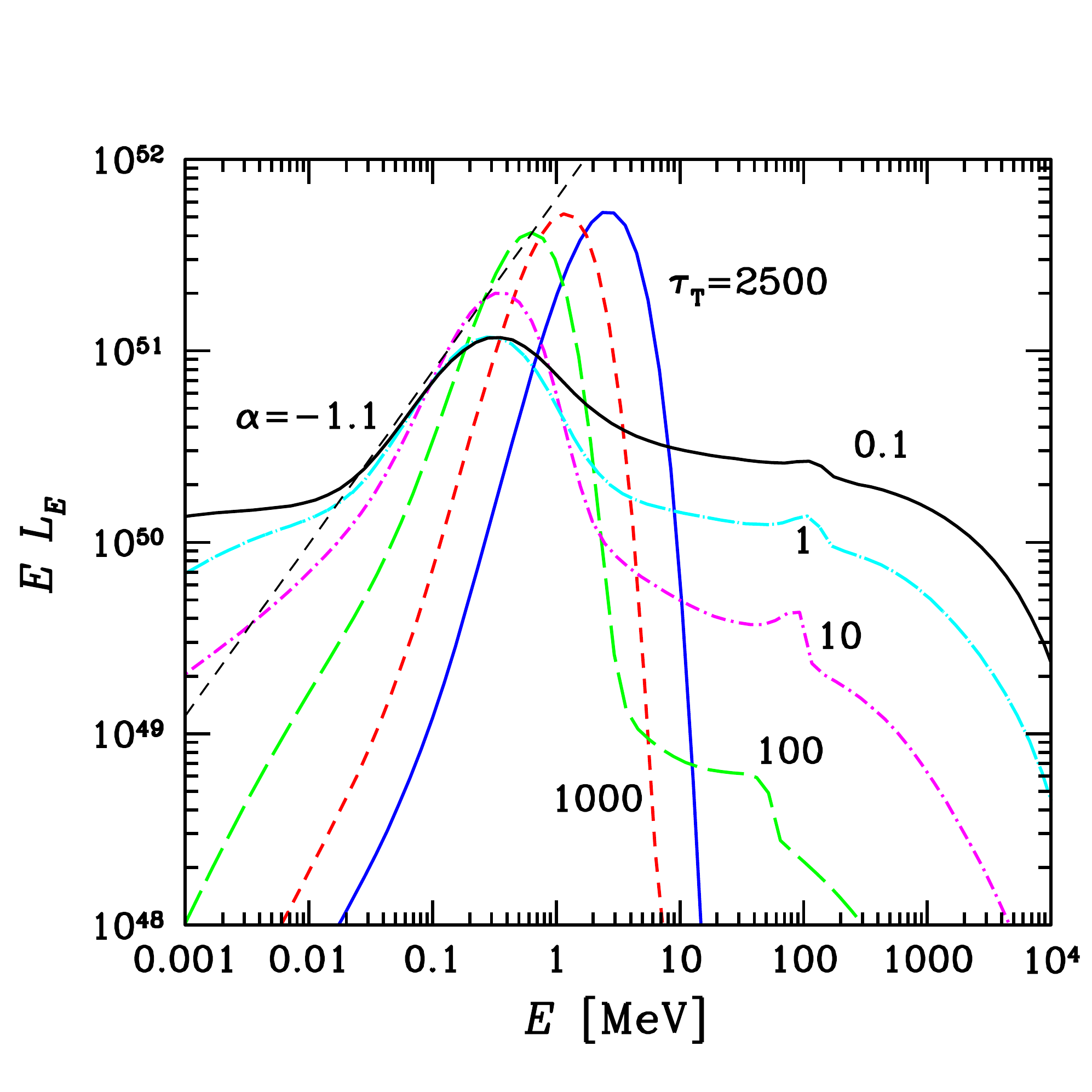}
\caption{Evolution of the radiation spectrum  
carried by the jet from high to low optical depths $\tauT$ in a sample model with
$L=2\times 10^{52}$ erg/s, $\eta=190$, and $\epsB=\sigma/2= 10^{-2}$.
The simulation started at radius $R_c=10^{11}$~cm with $\Gamma_c=20$,
and the jet was continually heated with $\epsilon_{\rm th}=\epsilon_{\rm nth}=0.025$.
The spectra are measured in the rest frame of the central engine and
not corrected for a cosmological redshift. 
The black curve ($\tauT=0.1$) shows the observed (escaping) spectrum that 
result from this evolution. (From \citet{Vurm+16grbphot}.)
}
\label{fig:sp_evol}       
\end{figure*}

(1) Generation of photons below the Wien radius ($\tauT\simgt 10^2$)
and their Comptonization to the Wien peak. 
A major source of photons is the synchrotron emission from nonthermal leptons.
If the magnetization is relatively weak, e.g. $\epsB\sim 10^{-2}$, then the synchrotron 
power is modest, however the {\it number} of generated photons 
is substantial, because they are emitted with low energies.
Below the Wien radius $\rW$ many of them are Comptonized to the Wien peak and 
eventually dominate the peak, significantly increasing its photon number and shifting
$\Epk$ to lower energies. 
The decrease of $\Epk$ due to the continuing photon supply to the peak ends 
when $\tauT$ is reduced below $\sim 10^2$, which approximately corresponds to $\rW$.

(2) Broadening of the spectrum by unsaturated Comptonization
outside the Wien radius, leading to a nonthermal shape of the spectrum.
The thermal Comptonization gradually switches to the unsaturated regime at $r\sim\RW$
and then the Compton $y$-parameter stays near unity.
The production of synchrotron photons continues at $r>\rW$ and
results in the soft ``excess'' seen below a few$\times 10$~keV in 
the emitted spectrum (Figure~\ref{fig:sp_evol}).
More importantly, unsaturated Comptonization of these photons plays 
a key role  in determining the spectral slope $\alpha$ below the peak.
The slope begins to soften near the Wien radius and attains its final value $\alpha\sim -1$
near the photosphere. The value of $\alpha$ depends on the heating history 
at $r>\rW$, in particular on the nonthermal dissipation channel, and is expected
to vary, however $\alpha\sim -1$ is a characteristic signature of unsaturated 
Comptonization. 

The high-energy tail of the spectrum develops mostly near the photosphere,
due to two effects, thermal and nonthermal.
Outside the Wien radius the electron temperature rises 
significantly above $\Epk/\Gamma$ and thermal Comptonization 
begins to produce photons with energies $E>\Epk$.
As the optical depth drops, the nonthermal high-energy 
component becomes increasingly prominent, especially in weakly magnetized jets
with the dominant IC cooling. Then the overlapping thermal and nonthermal 
Comptonization components together form an extended high-energy spectrum, 
which may superficially appear as a single emission component.

As seen in Figure~\ref{fig:sp_evol},
the spectrum has a distinctly nonthermal shape already
well below the photosphere (at $\tauT\simgt 10$).
In particular, the low-energy slope is significantly softer than
the Planck (or Wien) spectrum. Even if dissipation stopped completely at 
$\tauT\sim 10$, a thermal-looking prompt emission would not be expected;
instead, the emerging spectrum would resemble an exponentially
cutoff power law.

\begin{figure*}[t]
\vspace*{-1cm}
\includegraphics[width=0.77\textwidth]{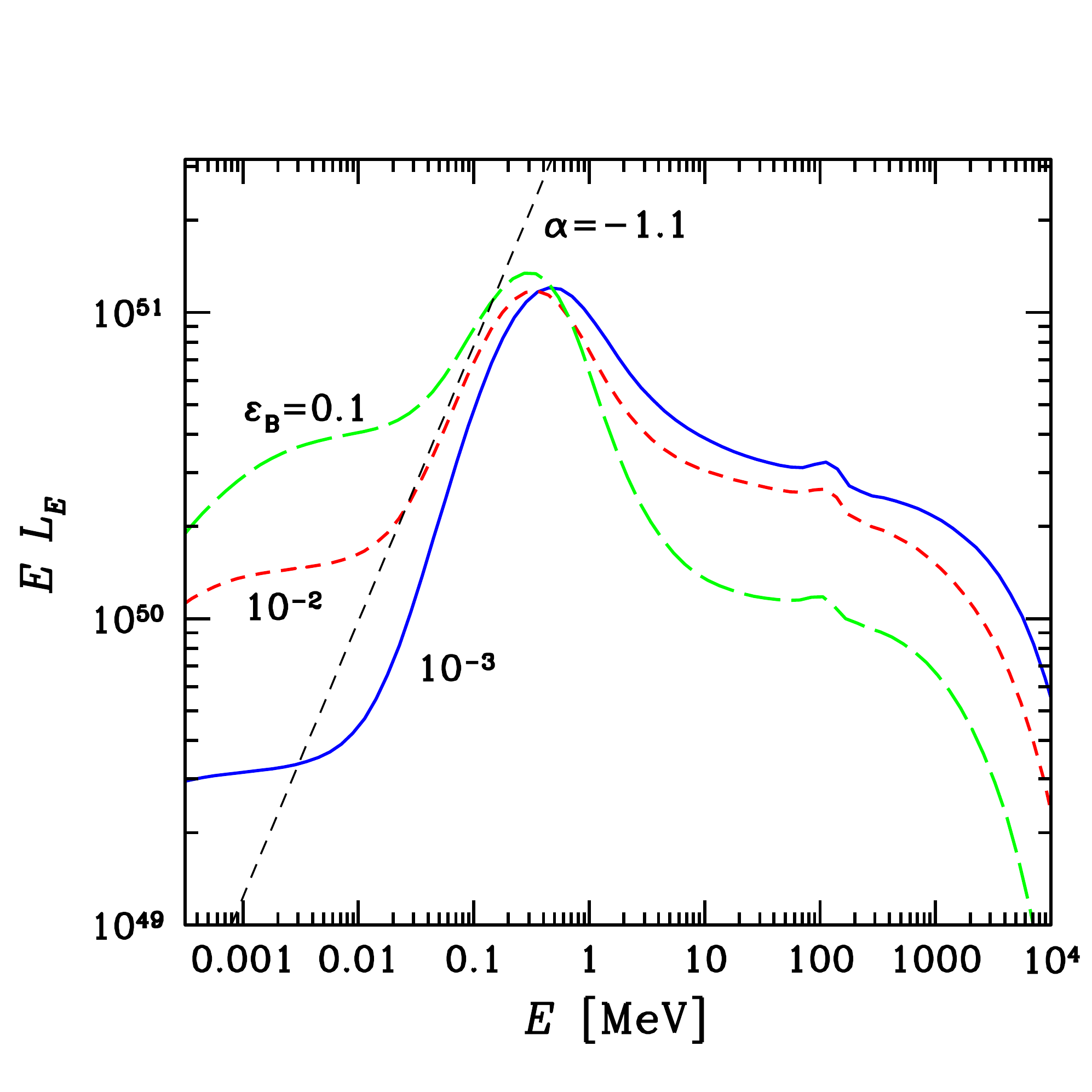}
\caption{Effect of varying magnetization 
on the emerging spectrum \citep{Vurm+16grbphot}. If variations of $\epsB$ are 
not resolved, a superposition of the spectra will be observed. 
}
\label{fig:sp_epsB}       
\end{figure*}

The dependence of the photospheric spectrum on the jet magnetization $\epsB$
is shown in Figure~\ref{fig:sp_epsB}.
Very weak magnetization $\epsB<10^{-3}$ is disfavored for typical GRBs,
as it leads to a hard slope $\alpha$ at low energies 
--- then practically the entire spectrum is dominated by the Comptonized
photons advected from very large optical depths, with little photon production at 
moderate $\tauT$.
The preferred magnetization is in the range $10^{-3}<\epsB<0.1$,
which gives $\alpha$ varying around $-1$.
At the low-$\epsB$ end, the nonthermal $e^\pm$ cascades are particularly efficient,}
and the pairs outnumber protons on average by a factor of $\simgt 10$. 

A large fraction of produced pairs ``freeze out'' in the expanding jet, so the jet remains
forever dominated by $e^\pm$ pairs. This should affect the afterglow emission 
produced by the reverse shock when the jet interacts with the external medium. 
Thus, the reverse-shock component in a GRB afterglow could serve as a probe of 
pair loading.

In strongly magnetized jets with $\epsB\simgt 1$, the pair cascade and
nonthermal Comptonization are suppressed, leading to a different radiation spectrum 
at high energies \citep{Vurm+11phot}. Radiation from magnetically dominated
jets was recently considered by  \cite{Gill+14pairphot,Thompson+14phot} and
\cite{Begue+15magphot}.
Gill and Thompson envision a two-stage evolution of 
a jet that starts out baryon-free and dominated by the Poynting flux. In their scenario,
baryon loading occurs at a large radius, and
the jet opacity is dominated by electron-positron 
pairs generated by dissipation, which takes place in two separate episodes.
Despite the complicated details, the evolution of radiative processes in their picture 
resembles that shown in Figure~\ref{fig:sp_evol}
--- basically, generation of synchrotron photons at large optical depths is
followed by Comptonization into a Band-like spectrum.

\subsection{Comparison with observed spectra}

A continuously heated and moderately magnetized jet generates a 
Band-type photospheric spectrum, with spectral slopes $\alpha$, $\beta$, and 
peak position $\Epk$ consistent with observations.
The radiative transfer simulations show a moderate dependence of $\Epk$ on 
the parameters of the problem, and no fine-tuning is needed to bring the peak 
into the observed range around 1~MeV.

A typical burst with $L\approx 10^{52}$ erg/s, a canonical Band spectrum, and 
$\Epk\simlt 1$~MeV is reproduced by the model if: 
\begin{enumerate}
\item 
The jet magnetization is in the range $10^{-3}<\epsB< 0.1$.
Very weak magnetization increases $\Epk$ by suppressing synchrotron emission;
strong magnetization softens the spectrum both below and above the peak,
and generates a prominent soft ``excess'' below a few tens of keV.
\item 
The jet Lorentz factor $\Gamma(\rcoll)=10-100$
at radii comparable to that of the stellar progenitor. Jets with a low $\Gamma(\rcoll)$ 
are more ``photon-rich'' and produce GRBs with low $\Ep$.
\item 
Heating has a nonthermal component that injects relativistic leptons into the jet.
The subphotopsheric nonthermal particles play a key role by providing soft
synchrotron photons as seeds for Comptonization. This impacts the 
formation of both $\Ep$ (at $\tauT\simgt 10^2$) and slope $\alpha$ (at $1<\tau<10^2$).
The absence of nonthermal leptons would lead to hard spectra with high $\Epk$.
\end{enumerate}

The detailed radiative transfer models have been applied to three well-studied
bright bursts, GRB 990123, GRB 090902B, and GRB 130427A,
which show different prompt spectra, and successful fits have been found in all 
three cases \citep{Vurm+16grbphot}.
The fits gave estimates for the main parameters of the GRB jets (Table~1).
In particular, the best-fit magnetization varies between 0.01 and 0.1, and the 
final Lorentz factor $\Gamma_f$ between 300 and 1200.
The corresponding photospheric radius $\Rph$ varies around $10^{13}$~cm.

\begin{table}[t]
\caption{Jet parameters for three fitted GRBs (from \cite{Vurm+16grbphot}).
}
\label{tab:1}       
\begin{tabular}{lcccccc}
\hline\noalign{\smallskip}
GRB  & $L_{54}$ & $\Gamma_c$ & $\Gamma_f$ & $\epsB$  & $\eps_{0,\rm th}$ & 
$\eps_{0,\rm nth}$ \\ 
\noalign{\smallskip}\hline\noalign{\smallskip}
 990123  	& 0.44 &  35  &  500  & 0.018  & 0.099  &  0.021  \\ 
 090902B 	& 2.3   &  70  & 1220 & 0.012  & 0.055  &  0.065  \\ 
 130427A 	& 0.47 & 64   &   300 & 0.046  & 0.093  &  0.067  \\ 
\noalign{\smallskip}\hline
\end{tabular}
\end{table}

The obtained $\Gamma_f$ can be compared  with the results of another, 
independent method of estimating the jet Lorentz factor. This method is
based on the reconstruction of the GeV+optical flash 
produced by the jet at much larger radii 
$r\simgt 10^{16}$~cm where it drives the external blast wave
\citep{Beloborodov+14pairgev,Vurm+14pairgevopt,Hascoet+15grblf}.
Remarkably, the prompt emission fit for GRB 130427A gave 
$\Gamma_{\rm f} \approx 300$, close to
the ejecta Lorentz factor $\Gamma_{\rm ej} \approx 350$
obtained from the GeV+optical flash reconstruction (Vurm et al. 2014). 
The fit to GRB~090902B gave $\Gamma_{\rm f} \approx 1200$,
which is higher than $\Gamma_{\rm ej} = 600-900$ used to fit the GeV flash.
However, in this case there is no significant inconsistency, because
the reverse shock in the blast wave from GRB~090902B is 
relativistic; in this case the flash modeling only gives
a lower limit  $\Gamma_{\rm ej} > 600$ \citep{Hascoet+15grblf}. 
A similar $\Gamma\approx 1000$ was proposed by \cite{Peer+12-090902b} 
based on a different phenomenological model for GRB~090902B (a multicolor blackbody 
for the photosphere and nonthermal radiation from dissipation at a large radius).

Besides the values of $\Gamma$ and $\epsB$, the radiative transfer models 
of the prompt emission give insights into the heating history of the GRB jets.
The fit to GRB~130427A requires that the nonthermal heating be significantly deep 
below the photosphere and become
weak well before the jet expands to transparency; 
the observed Band-like spectrum is mainly shaped by thermal Comptonization. 
In GRB~990123, thermal heating appears to dominate at all radii, and
the nonthermal synchrotron source is relatively weak.
The model still predicts a moderate excess below a few tens of keV due to 
synchrotron emission. A hint of such excess is indeed seen in the data (Figure 2 in 
\cite{Briggs+99-990123}; R. Preece, private communication).

\begin{figure*}[t]
\vspace*{-1cm}
  \includegraphics[width=0.77\textwidth]{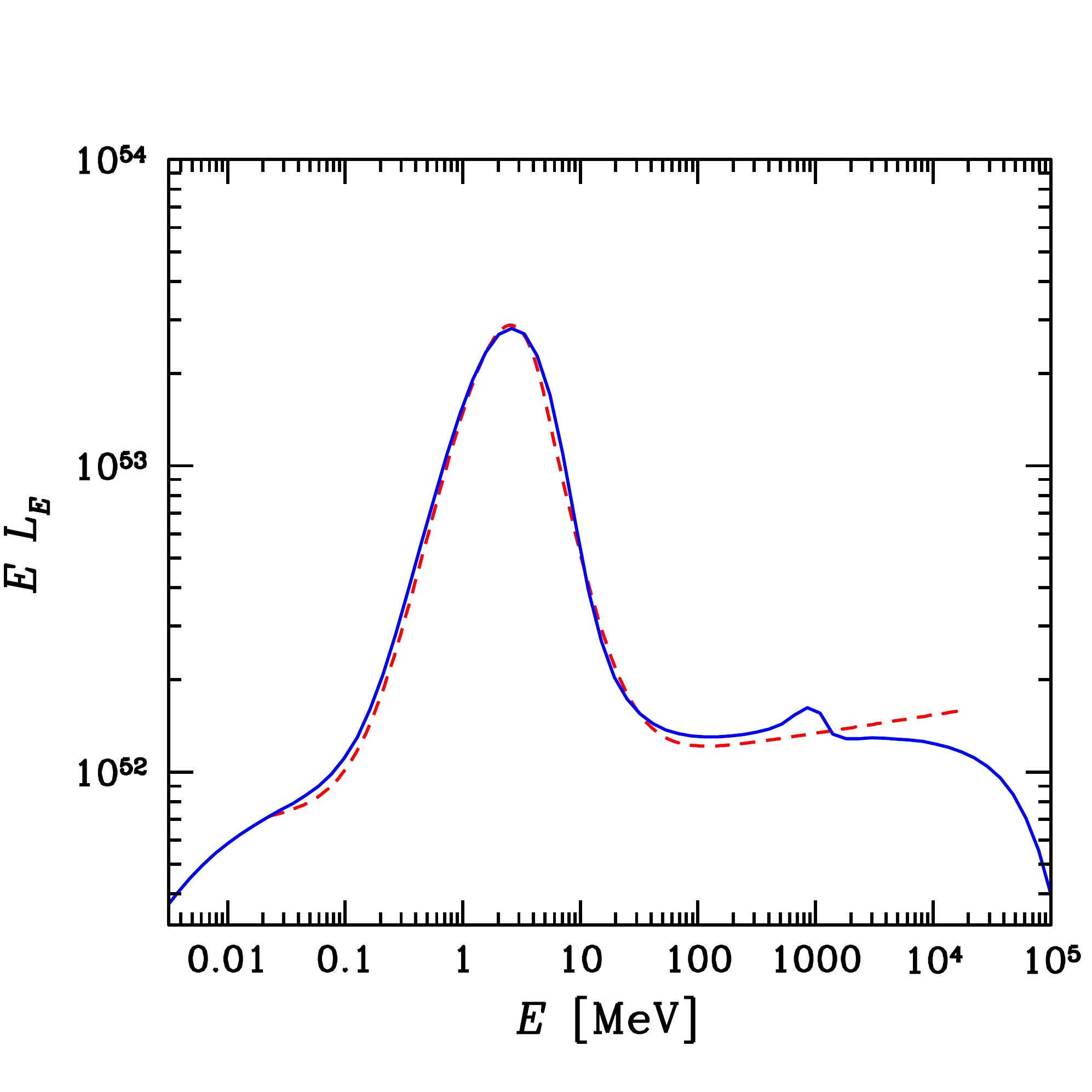}
\caption{Radiative transfer model for GRB 090902B (solid curve). The dashed curve shows 
the Band$+$power-law fit by \cite{Abdo+09-090902}, in time bin b.
(From \cite{Vurm+16grbphot}).}
\label{fig:090902B}       
\end{figure*}

In contrast, nonthermal dissipation in GRB~090902B is found to be strong up to the 
photosphere and beyond. It explains the observed high-energy component 
and the soft excess, which were previously modeled as a power law of unknown origin 
\citep{Abdo+09-090902}. The two models are compared in Figure~\ref{fig:090902B}.
The radiative transfer model shows that the soft excess is the synchrotron emission
and the high-energy component is IC emission from nonthermal electrons.\footnote{A
different hadronic shock model explanation was also proposed by \cite{Asano+10optex}}
They are not parts of a single power law, however they are produced by the same
nonthermal electron population.
The high-energy component of the prompt emission is also expected to contaminate 
(at early times) the GeV flash observed in GRB~090902B by {\it Fermi} LAT.


\section{Polarization}

Linear polarization measurements  in the hard X-ray band have been reported for GRB~041219A \citep{KalEtAl:2007, McGEtAl:2007, GotEtAl:2009}, GRB~061122 \citep{McGEtAl:2009, GotEtAl:2013}, GRB~100826A \citep{YonEtAl:2011}, GRB~110301A, GRB~110721A \citep{YonEtAl:2012}, and GRB~140206A \citep{GotEtAl:2014}. However, 
the detection significance for all these bursts is modest, because of a low 
sensitivity, and the measurements have poorly known systematic uncertainties.
Future detectors can significantly improve the polarization information, which
will be highly valuable for GRB modeling.

Radiative transfer in the jet occurs in two linear polarization states with different intensities, 
however this gives no observable polarization as long as the emission is 
symmetric about the line of sight (Beloborodov 2011). Due to the radial Doppler 
beaming of radiation, the observed GRB radiation is 
expected to come from a narrow region around the line of sight, of angular size 
$\delta\theta\sim\Gamma^{-1}$. Its symmetry can be broken in two ways.

(1) A preferred direction for polarization appears if there is a strong transverse gradient in 
the jet parameters at the photosphere (e.g. a gradient in luminosity, Lorentz factor, etc.). 
Then observable polarization emerges after the last scattering around $\Rph$, which 
was investigated by \citet{Lundman+14grbpol} and \citet{ItoEtAl:2014}. 
They predicted polarization degrees up to $\Pi \sim 40 \%$ if the jet has 
significant structure on angular scales $\delta\theta\sim\Gamma^{-1}$. In particular, 
a strong transverse gradient is expected near the jet edge $\theta=\theta_c$.
If the collimation angle $\theta_c$ 
is not much larger than $\Gamma^{-1}$ then observers detecting a bright burst (but not exactly on the 
jet axis) will see the jet 
cross section $\sim\pi r^2\theta_c^2$ at some inclination.
Then the line of sight and the jet axis form a preferred plane and
a strong linear polarization should show up in the main peak of the GRB spectrum.
\citet{Lundman+14grbpol} also showed that for a jet with $\theta_c\gg\Gamma^{-1}$ the 
probability to observe its edge (and hence to break the symmetry of observed emission)
is $\sim 4(\Gamma\theta_c)^{-1}$.
For instance, if $\theta_c\sim 0.1$ and $\Gamma \approx 400$ the edge should
be visible in roughly 10\% of detected GRBs.
The actual distribution of $\theta_c$ is however unknown;
attempts to infer this distribution from the analysis of so-called ``jet breaks'' in afterglow 
light curves have been inconclusive.

(2) If part of the GRB spectrum is dominated by synchrotron radiation, a preferred 
direction for polarization can be set by the magnetic field where this radiation is produced.
The transfer models discussed in \Sect~4 show that the main MeV peak is dominated
by the Comptonized radiation advected from large optical depths. The Comptonized 
peak is unpolarized as long as there is no significant structure on angular scales 
$\delta\theta\sim\Gamma^{-1}$. However, at sufficiently low energies $E\ll\Epk$ the 
spectrum is dominated by synchrotron emission.
This leads to a unique polarization signature: a rise in GRB polarization 
toward lower energies (Lundman et al. 2016).

Synchrotron emission from relativistic electrons in a uniform magnetic field 
${\mathbf B}$ is linearly polarized in the plane perpendicular to ${\mathbf B}$. 
The polarization degree for an isotropic electron distribution is 
$\Pi_{\rm syn}=(p+1)/(p+7/3)$, 
where $p=d\ln N/d\ln E_e$ is the slope of the electron spectrum 
(e.g. \citealt{Rybicki_book}). This standard result is somewhat modified when the 
observed region is a spherical patch in a relativistic outflow, carrying an ordered 
transverse magnetic field, which gives $\Pi_{\rm syn}$ varying around 50\%, 
depending on $p$ \citep{LyuParBla:2003}. Synchrotron radiation produced well 
below the photosphere will lose its polarization after a few scatterings, however
a significant fraction of synchrotron photons produced around and above 
the photosphere will escape without scattering and preserve their polarization. 
The observed GRB spectrum $L(E)$ can be viewed as the sum of
two contributions: photons escaping the jet after their last Compton scattering, 
$L_{\rm sc}(E)$, and photons escaping directly after their emission by the synchrotron 
mechanism, with no scattering, $L_{\rm nsc}(E)$. The observed polarization is then 
given by
\beq
\label{eq:f}
  \Pi(E) = f_{\rm nsc}(E)\Pi_{\rm syn}, \qquad f_{\rm nsc}(E)=\frac{L_{\rm nsc}(E)}{L(E)}.
\eeq

\begin{figure*}[t]
\hspace*{0.4cm}
  \includegraphics[width=0.88\textwidth]{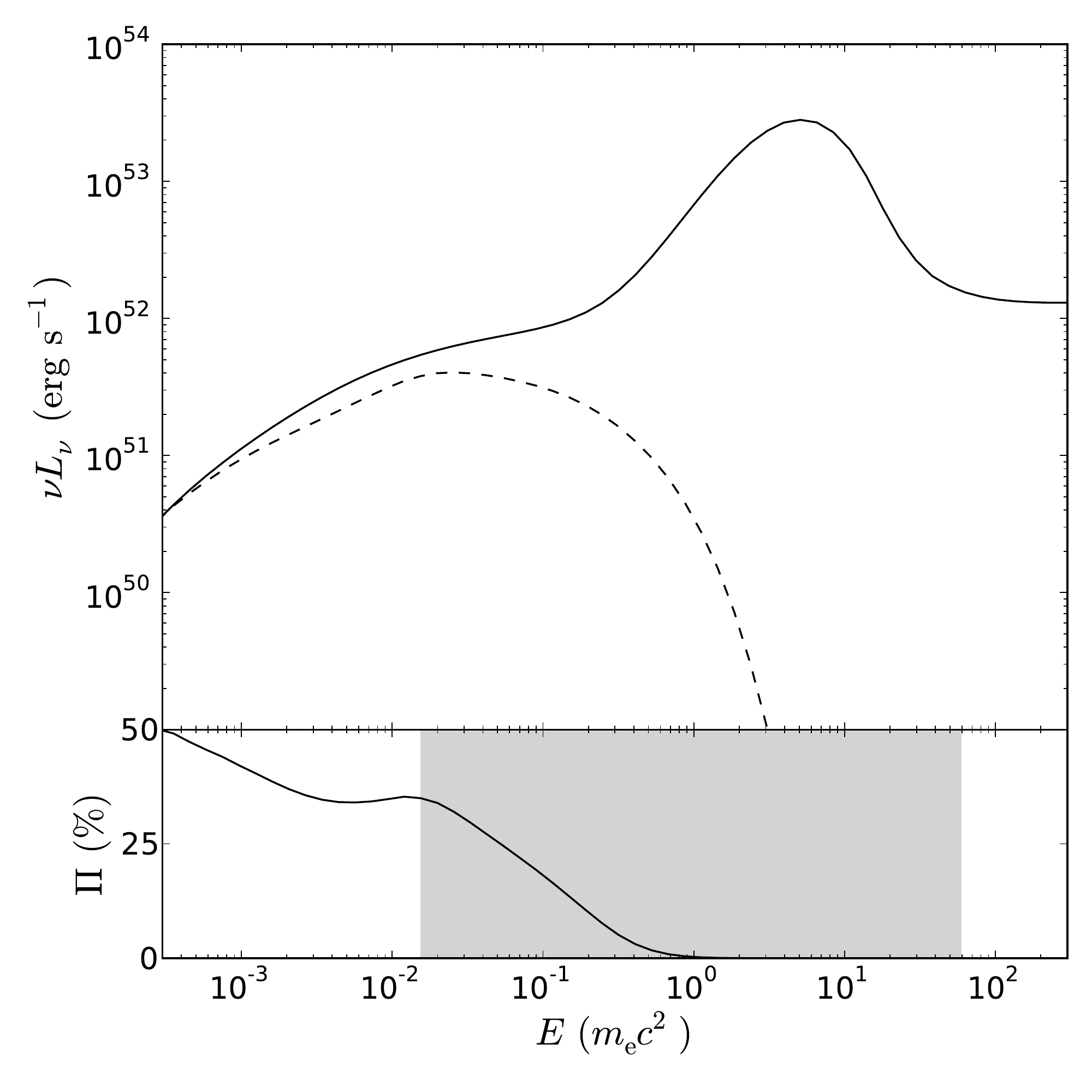}
\caption{Top panel: the simulated spectrum of GRB~090902B (solid curve) and 
the contribution of unscattered synchrotron emission (dashed curve).
Their ratio gives the unscattered fraction $f_{\rm nsc}(E)$.
Bottom panel: expected polarization as a function of photon energy $E$. 
The shaded region shows the Fermi GBM band (NaI + BGO detectors, 8 keV to 30 MeV).
(From \citet{Lundman+16}.)}
\label{fig:pol}       
\end{figure*}

Using the transfer models of \citet{Vurm+16grbphot}, Lundman et al. (2016)
calculated $f_{\rm nsc}(E)$ and the expected polarization degree $\Pi(E)$ for 
GRBs 990123, 090902B, and 110721A. Their results demonstrate that 
strong polarization is expected in the X-ray band under two conditions:
(1) significant dissipation extends to $r\simgt \Rph$ and is accompanied by injection of
nonthermal particles, and (2) the jet magnetization is significant, $\epsB\simgt 10^{-2}$.
These conditions are well satisfied in GRB~090902B (\Sect~4).
Then the GRB spectrum has a prominent soft excess emerging above the Band spectrum
at low energies, and
a significant fraction of this excess is formed by unscattered synchrotron photons.
This leads to the strong polarization rise toward low energies (Figure~\ref{fig:pol}).

\section{Possible future developments}

The detailed radiative transfer simulations can be systematically applied to all available 
spectra of prompt GRB emission, including those resolved in time.
Such fits would reconstruct the jet heating and magnetization 
for a large sample of GRBs and may reveal important correlations. This could shed light 
on how the jets are produced and heated, and what role is played in this process by the 
magnetic field.

The analysis of prompt emission can be combined with that  
of GRB afterglow, especially its early phase. The GeV+optical flash predicted and observed 
at the onset of the afterglow \citep{Beloborodov+14pairgev}
provides particularly useful data that recently helped to disentangle the main parameters 
of the blast wave in seven GRBs \citep{Hascoet+15grblf}. The initial Lorentz factor of the blast 
wave reconstructed by this method can be compared with the jet Lorentz factor inferred
from the transfer simulations of the prompt emission, as has been done for GRB~130427A.
An additional test would be provided by future detections of the predicted TeV counterpart
of the GeV+optical flash, which should rise with some delay \citep{Hascoet+15grblf}.
It is detectable by current ground-based Cherenkov telescopes and should
certainly be detected by the future Cherenkov Telescope Array (CTA) \citep{Vurm+TeV16}.

Spectra of prompt GRB emission are poorly known at low energies, from soft X-rays to 
the optical band. While the main MeV peak  is dominated by the Comptonized radiation
advected from large optical depths, the prompt optical radiation must be dominated 
by synchrotron emission outside the scattering photosphere. The low-frequency emission 
is predicted to be affected by synchrotron self-absorption, which tends to give
a flat spectral slope \citep{Vurm+11phot}. While prompt optical emission has been detected
in some GRBs, there is currently no information on its spectral slope. A color measurement 
and better sampling of the optical light curve would provide useful information about 
dissipation outside the photosphere and inside of the external blast wave. This is a rather
broad range of radii, even taking into account that the photosphere can be ``carried'' 
forward by the $e^\pm$-dressed internal shocks up to two decades in radius (\Sect~2).
Radiation produced outside the photosphere is expected to have a broad nonthermal spectrum
with no sharp MeV peak and a slower variability on timescales $\sim r/c\Gamma^2$.

Important developments may result from future observations of polarization of the 
prompt GRB emission and its variation across the burst spectrum. An unpolarized MeV 
peak and a rise in 
linear polarization toward low photon energies would confirm the synchrotron contribution 
at $E\ll\Epk$ predicted by the radiative transfer simulations (\Sects~4 and 5). Bursts similar 
to GRB~090902B are the most promising targets for such observations. Observations of 
polarization of the main MeV peak of photospheric emission would provide evidence for 
a transverse structure of the jet on angular scales $\delta\theta\sim\Gamma^{-1}$.

Detection of neutrinos from GRB explosions is possible in the near future. 
The detection (or strong upper limits) would test the ideas 
for the dissipation mechanism in the relativistic jet, including that in the opaque region well 
below the photosphere. In particular, the expected presence of a free neutron component 
leads to strong nuclear collisional dissipation, which produces 
10-100~GeV neutrinos detectable by IceCube DeepCore.

\begin{acknowledgements}
AMB's research is supported by NSF grant AST-1412485, NASA grant NNX15AE26G,
and a grant from the Simons Foundation (\#446228, Andrei Beloborodov). 
PM's research is supported by NASA grant NNX13AH50G.

\end{acknowledgements}



\bibliographystyle{aps-nameyear}      

\bibliography{rev_final.bbl}                

\begin{thebibliography}{93}
\ifx \bisbn   \undefined \def \bisbn  #1{ISBN #1}\fi
\ifx \binits  \undefined \def \binits#1{#1} \fi
\ifx \bauthor  \undefined \def \bauthor#1{#1} \fi
\ifx \bjtitle  \undefined \def \bjtitle#1{\textrm{#1}}\fi
\ifx \batitle  \undefined \def \batitle#1{#1} \fi
\ifx \bctitle  \undefined \def \bctitle#1{#1} \fi
\ifx \bvolume  \undefined \def \bvolume#1{\textbf{#1}}\fi
\ifx \byear  \undefined \def \byear#1{#1} \fi
\ifx \bissue  \undefined \def \bissue#1{#1} \fi
\ifx \bfpage  \undefined \def \bfpage#1{#1} \fi
\ifx \blpage  \undefined \def \blpage #1{#1} \fi
\ifx \burl  \undefined \def \burl#1{#1} \fi
\ifx \doiurl  \undefined \def \doiurl#1{#1} \fi
\ifx \betal  \undefined \def \betal{et al.} \fi
\ifx \binstitute  \undefined \def \binstitute#1{#1} \fi
\ifx \beditor  \undefined \def \beditor#1{#1} \fi
\ifx \bpublisher  \undefined \def \bpublisher#1{#1} \fi
\ifx \bbtitle  \undefined \def \bbtitle#1{\textit{#1}} \fi
\ifx \bedition  \undefined \def \bedition#1{#1} \fi
\ifx \bseriesno  \undefined \def \bseriesno#1{#1} \fi
\ifx \blocation  \undefined \def \blocation#1{#1} \fi
\ifx \bsertitle  \undefined \def \bsertitle#1{#1} \fi
\ifx \bsnm \undefined \def \bsnm#1{#1} \fi
\ifx \bsuffix \undefined \def \bsuffix#1{#1} \fi
\ifx \bparticle \undefined \def \bparticle#1{#1} \fi
\ifx \barticle \undefined \def \barticle#1{#1} \fi
\ifx \botherref \undefined \def \botherref #1{#1} \fi
\ifx \url \undefined \def \url#1{#1} \fi
\ifx \bchapter \undefined \def \bchapter#1{#1} \fi
\ifx \bbook \undefined \def \bbook#1{#1} \fi
\ifx \bcomment \undefined \def \bcomment#1{#1} \fi
\ifx \oauthor \undefined \def \oauthor#1{#1} \fi
\ifx \citeauthoryear \undefined \def \citeauthoryear#1{#1} \fi
\ifx \texttildelow  \undefined \def \texttildelow{\symbol{126}} \fi
\def \endbibitem {}
\ifx \bconflocation  \undefined \def \bconflocation#1{#1} \fi

\bibitem[\protect\citeauthoryear{{Abdo} and the
  {Fermi}~collaboration}{2009}]{Abdo+09-090902}
\begin{barticle}
\bauthor{\binits{A.A.} \bsnm{{Abdo}}},
\bauthor{\bparticle{the} \bsnm{{Fermi}~collaboration}},
\batitle{{Fermi Observations of GRB 090902B: A Distinct Spectral Component in
  the Prompt and Delayed Emission}}.
\bjtitle{\apjl}
\bvolume{706},
\bfpage{138}--\blpage{144}
(\byear{2009}).
doi:\doiurl{10.1088/0004-637X/706/1/L138}
\end{barticle}
\endbibitem

\bibitem[\protect\citeauthoryear{{Ackermann} and {The Fermi
  collaboration}}{2013}]{Ackermann+13-110731}
\begin{barticle}
\bauthor{\binits{M.} \bsnm{{Ackermann}}},
\bauthor{\bsnm{{The Fermi collaboration}}},
\batitle{{Multiwavelength Observations of GRB 110731A: GeV Emission from Onset
  to Afterglow}}.
\bjtitle{\apj}
\bvolume{763},
\bfpage{71}
(\byear{2013}).
doi:\doiurl{10.1088/0004-637X/763/2/71}
\end{barticle}
\endbibitem

\bibitem[\protect\citeauthoryear{{Asano} et~al.}{2010}]{Asano+10optex}
\begin{barticle}
\bauthor{\binits{K.} \bsnm{{Asano}}},
\bauthor{\binits{S.} \bsnm{{Inoue}}},
\bauthor{\binits{P.} \bsnm{{M{\'e}sz{\'a}ros}}},
\batitle{{Prompt X-ray and Optical Excess Emission Due to Hadronic Cascades in
  Gamma-ray Bursts}}.
\bjtitle{\apjl}
\bvolume{725},
\bfpage{121}--\blpage{125}
(\byear{2010}).
doi:\doiurl{10.1088/2041-8205/725/2/L121}
\end{barticle}
\endbibitem

\bibitem[\protect\citeauthoryear{{Axelsson} and
  {Borgonovo}}{2015}]{Axelsson+15spwidth}
\begin{barticle}
\bauthor{\binits{M.} \bsnm{{Axelsson}}},
\bauthor{\binits{L.} \bsnm{{Borgonovo}}},
\batitle{{The width of gamma-ray burst spectra}}.
\bjtitle{\mnras}
\bvolume{447},
\bfpage{3150}--\blpage{3154}
(\byear{2015}).
doi:\doiurl{10.1093/mnras/stu2675}
\end{barticle}
\endbibitem

\bibitem[\protect\citeauthoryear{{Bahcall} and
  {M{\'e}sz{\'a}ros}}{2000}]{Bahcall+00pn}
\begin{barticle}
\bauthor{\binits{J.N.} \bsnm{{Bahcall}}},
\bauthor{\binits{P.} \bsnm{{M{\'e}sz{\'a}ros}}},
\batitle{{5-10 GeV Neutrinos from Gamma-Ray Burst Fireballs}}.
\bjtitle{Physical Review Letters}
\bvolume{85},
\bfpage{1362}--\blpage{1365}
(\byear{2000})
\end{barticle}
\endbibitem

\bibitem[\protect\citeauthoryear{{Band} et~al.}{2009}]{Band+09}
\begin{barticle}
\bauthor{\binits{D.L.} \bsnm{{Band}}},
\bauthor{\binits{M.} \bsnm{{Axelsson}}},
\bauthor{\binits{L.} \bsnm{{Baldini}}},
\bauthor{\binits{G.} \bsnm{{Barbiellini}}},
\bauthor{\binits{M.G.} \bsnm{{Baring}}},
\bauthor{\binits{D.} \bsnm{{Bastieri}}},
\bauthor{\binits{M.} \bsnm{{Battelino}}},
\bauthor{\binits{R.} \bsnm{{Bellazzini}}},
\bauthor{\binits{E.} \bsnm{{Bissaldi}}},
\bauthor{\binits{G.} \bsnm{{Bogaert}}},
\bauthor{\binits{J.} \bsnm{{Bonnell}}},
\bauthor{\binits{J.} \bsnm{{Chiang}}},
\bauthor{\binits{J.} \bsnm{{Cohen-Tanugi}}},
\bauthor{\binits{V.} \bsnm{{Connaughton}}},
\bauthor{\binits{S.} \bsnm{{Cutini}}},
\bauthor{\binits{F.} \bsnm{{de Palma}}},
\bauthor{\binits{B.L.} \bsnm{{Dingus}}},
\bauthor{\binits{E.} \bsnm{{do Couto e Silva}}},
\bauthor{\binits{G.} \bsnm{{Fishman}}},
\bauthor{\binits{A.} \bsnm{{Galli}}},
\bauthor{\binits{N.} \bsnm{{Gehrels}}},
\bauthor{\binits{N.} \bsnm{{Giglietto}}},
\bauthor{\binits{J.} \bsnm{{Granot}}},
\bauthor{\binits{S.} \bsnm{{Guiriec}}},
\bauthor{\binits{R.E.} \bsnm{{Hughes}}},
\bauthor{\binits{T.} \bsnm{{Kamae}}},
\bauthor{\binits{N.} \bsnm{{Komin}}},
\bauthor{\binits{F.} \bsnm{{Kuehn}}},
\bauthor{\binits{M.} \bsnm{{Kuss}}},
\bauthor{\binits{F.} \bsnm{{Longo}}},
\bauthor{\binits{P.} \bsnm{{Lubrano}}},
\bauthor{\binits{R.M.} \bsnm{{Kippen}}},
\bauthor{\binits{M.N.} \bsnm{{Mazziotta}}},
\bauthor{\binits{J.E.} \bsnm{{McEnery}}},
\bauthor{\binits{S.} \bsnm{{McGlynn}}},
\bauthor{\binits{E.} \bsnm{{Moretti}}},
\bauthor{\binits{T.} \bsnm{{Nakamori}}},
\bauthor{\binits{J.P.} \bsnm{{Norris}}},
\bauthor{\binits{M.} \bsnm{{Ohno}}},
\bauthor{\binits{M.} \bsnm{{Olivo}}},
\bauthor{\binits{N.} \bsnm{{Omodei}}},
\bauthor{\binits{V.} \bsnm{{Pelassa}}},
\bauthor{\binits{F.} \bsnm{{Piron}}},
\bauthor{\binits{R.} \bsnm{{Preece}}},
\bauthor{\binits{M.} \bsnm{{Razzano}}},
\bauthor{\binits{J.J.} \bsnm{{Russell}}},
\bauthor{\binits{F.} \bsnm{{Ryde}}},
\bauthor{\binits{P.M.} \bsnm{{Saz Parkinson}}},
\bauthor{\binits{J.D.} \bsnm{{Scargle}}},
\bauthor{\binits{C.} \bsnm{{Sgr{\`o}}}},
\bauthor{\binits{T.} \bsnm{{Shimokawabe}}},
\bauthor{\binits{P.D.} \bsnm{{Smith}}},
\bauthor{\binits{G.} \bsnm{{Spandre}}},
\bauthor{\binits{P.} \bsnm{{Spinelli}}},
\bauthor{\binits{M.} \bsnm{{Stamatikos}}},
\bauthor{\binits{B.L.} \bsnm{{Winer}}},
\bauthor{\binits{R.} \bsnm{{Yamazaki}}},
\batitle{{Prospects for GRB Science with the Fermi Large Area Telescope}}.
\bjtitle{\apj}
\bvolume{701},
\bfpage{1673}--\blpage{1694}
(\byear{2009}).
doi:\doiurl{10.1088/0004-637X/701/2/1673}
\end{barticle}
\endbibitem

\bibitem[\protect\citeauthoryear{{Bartos} et~al.}{2013}]{Bartos+13pn}
\begin{barticle}
\bauthor{\binits{I.} \bsnm{{Bartos}}},
\bauthor{\binits{A.M.} \bsnm{{Beloborodov}}},
\bauthor{\binits{K.} \bsnm{{Hurley}}},
\bauthor{\binits{S.} \bsnm{{M{\'a}rka}}},
\batitle{{Detection Prospects for GeV Neutrinos from Collisionally Heated
  Gamma-ray Bursts with IceCube/DeepCore}}.
\bjtitle{Physical Review Letters}
\bvolume{110}(\bissue{24}),
\bfpage{241101}
(\byear{2013}).
doi:\doiurl{10.1103/PhysRevLett.110.241101}
\end{barticle}
\endbibitem

\bibitem[\protect\citeauthoryear{{B{\'e}gu{\'e}} and
  {Pe'er}}{2015}]{Begue+15magphot}
\begin{barticle}
\bauthor{\binits{D.} \bsnm{{B{\'e}gu{\'e}}}},
\bauthor{\binits{A.} \bsnm{{Pe'er}}},
\batitle{{Poynting-flux-dominated Jets Challenged by their Photospheric
  Emission}}.
\bjtitle{\apj}
\bvolume{802},
\bfpage{134}
(\byear{2015}).
doi:\doiurl{10.1088/0004-637X/802/2/134}
\end{barticle}
\endbibitem

\bibitem[\protect\citeauthoryear{{Beloborodov}}{2003}]{Beloborodov03neutron}
\begin{barticle}
\bauthor{\binits{A.M.} \bsnm{{Beloborodov}}},
\batitle{{Nuclear Composition of Gamma-Ray Burst Fireballs}}.
\bjtitle{\apj}
\bvolume{588},
\bfpage{931}--\blpage{944}
(\byear{2003}).
doi:\doiurl{10.1086/374217}
\end{barticle}
\endbibitem

\bibitem[\protect\citeauthoryear{{Beloborodov}}{2010}]{Beloborodov10pn}
\begin{barticle}
\bauthor{\binits{A.M.} \bsnm{{Beloborodov}}},
\batitle{{Collisional mechanism for gamma-ray burst emission}}.
\bjtitle{\mnras}
\bvolume{407},
\bfpage{1033}--\blpage{1047}
(\byear{2010}).
doi:\doiurl{10.1111/j.1365-2966.2010.16770.x}
\end{barticle}
\endbibitem

\bibitem[\protect\citeauthoryear{{Beloborodov}}{2011}]{Beloborodov11radtran}
\begin{barticle}
\bauthor{\binits{A.M.} \bsnm{{Beloborodov}}},
\batitle{{Radiative Transfer in Ultrarelativistic Outflows}}.
\bjtitle{\apj}
\bvolume{737},
\bfpage{68}
(\byear{2011}).
doi:\doiurl{10.1088/0004-637X/737/2/68}
\end{barticle}
\endbibitem

\bibitem[\protect\citeauthoryear{{Beloborodov}}{2013}]{Beloborodov13peak}
\begin{barticle}
\bauthor{\binits{A.M.} \bsnm{{Beloborodov}}},
\batitle{{Regulation of the Spectral Peak in Gamma-Ray Bursts}}.
\bjtitle{\apj}
\bvolume{764},
\bfpage{157}
(\byear{2013}).
doi:\doiurl{10.1088/0004-637X/764/2/157}
\end{barticle}
\endbibitem

\bibitem[\protect\citeauthoryear{{Beloborodov}}{2016}]{Beloborodov16phot}
\begin{botherref}
\oauthor{\binits{A.M.} \bsnm{{Beloborodov}}},
{Sub-photospheric shocks in relativistic explosions}.
ArXiv e-prints
\textbf{1604.02794}
(2016)
\end{botherref}
\endbibitem

\bibitem[\protect\citeauthoryear{{Beloborodov}
  et~al.}{2014}]{Beloborodov+14pairgev}
\begin{barticle}
\bauthor{\binits{A.M.} \bsnm{{Beloborodov}}},
\bauthor{\binits{R.} \bsnm{{Hasco{\"e}t}}},
\bauthor{\binits{I.} \bsnm{{Vurm}}},
\batitle{{On the Origin of GeV Emission in Gamma-Ray Bursts}}.
\bjtitle{\apj}
\bvolume{788},
\bfpage{36}
(\byear{2014}).
doi:\doiurl{10.1088/0004-637X/788/1/36}
\end{barticle}
\endbibitem

\bibitem[\protect\citeauthoryear{{Beloborodov}
  et~al.}{2000}]{Beloborodov+00pdsgrb}
\begin{barticle}
\bauthor{\binits{A.M.} \bsnm{{Beloborodov}}},
\bauthor{\binits{B.E.} \bsnm{{Stern}}},
\bauthor{\binits{R.} \bsnm{{Svensson}}},
\batitle{{Power Density Spectra of Gamma-Ray Bursts}}.
\bjtitle{\apj}
\bvolume{535},
\bfpage{158}--\blpage{166}
(\byear{2000}).
doi:\doiurl{10.1086/308836}
\end{barticle}
\endbibitem

\bibitem[\protect\citeauthoryear{{Blandford} and
  {Payne}}{1981}]{Blandford+81radshock2}
\begin{barticle}
\bauthor{\binits{R.D.} \bsnm{{Blandford}}},
\bauthor{\binits{D.G.} \bsnm{{Payne}}},
\batitle{{Compton Scattering in a Converging Fluid Flow - Part Two - Radiation
  Dominated Shock}}.
\bjtitle{\mnras}
\bvolume{194},
\bfpage{1041}
(\byear{1981}).
doi:\doiurl{10.1093/mnras/194.4.1041}
\end{barticle}
\endbibitem

\bibitem[\protect\citeauthoryear{{Briggs} et~al.}{1999}]{Briggs+99-990123}
\begin{barticle}
\bauthor{\binits{M.S.} \bsnm{{Briggs}}},
\bauthor{\binits{D.L.} \bsnm{{Band}}},
\bauthor{\binits{R.M.} \bsnm{{Kippen}}},
\bauthor{\binits{R.D.} \bsnm{{Preece}}},
\bauthor{\binits{C.} \bsnm{{Kouveliotou}}},
\bauthor{\binits{J.} \bsnm{{van Paradijs}}},
\bauthor{\binits{G.H.} \bsnm{{Share}}},
\bauthor{\binits{R.J.} \bsnm{{Murphy}}},
\bauthor{\binits{S.M.} \bsnm{{Matz}}},
\bauthor{\binits{A.} \bsnm{{Connors}}},
\bauthor{\binits{C.} \bsnm{{Winkler}}},
\bauthor{\binits{M.L.} \bsnm{{McConnell}}},
\bauthor{\binits{J.M.} \bsnm{{Ryan}}},
\bauthor{\binits{O.R.} \bsnm{{Williams}}},
\bauthor{\binits{C.A.} \bsnm{{Young}}},
\bauthor{\binits{B.} \bsnm{{Dingus}}},
\bauthor{\binits{J.R.} \bsnm{{Catelli}}},
\bauthor{\binits{R.A.M.J.} \bsnm{{Wijers}}},
\batitle{{Observations of GRB 990123 by the Compton Gamma Ray Observatory}}.
\bjtitle{\apj}
\bvolume{524},
\bfpage{82}--\blpage{91}
(\byear{1999}).
doi:\doiurl{10.1086/307808}
\end{barticle}
\endbibitem

\bibitem[\protect\citeauthoryear{{Bromberg} and
  {Levinson}}{2009}]{Bromberg+09jet}
\begin{barticle}
\bauthor{\binits{O.} \bsnm{{Bromberg}}},
\bauthor{\binits{A.} \bsnm{{Levinson}}},
\batitle{{Recollimation and Radiative Focusing of Relativistic Jets:
  Applications to Blazars and M87}}.
\bjtitle{\apj}
\bvolume{699},
\bfpage{1274}--\blpage{1280}
(\byear{2009}).
doi:\doiurl{10.1088/0004-637X/699/2/1274}
\end{barticle}
\endbibitem

\bibitem[\protect\citeauthoryear{{Bromberg} et~al.}{2015}]{Bromberg+15jetcomp}
\begin{barticle}
\bauthor{\binits{O.} \bsnm{{Bromberg}}},
\bauthor{\binits{J.} \bsnm{{Granot}}},
\bauthor{\binits{T.} \bsnm{{Piran}}},
\batitle{{On the composition of GRBs' Collapsar jets}}.
\bjtitle{\mnras}
\bvolume{450},
\bfpage{1077}--\blpage{1084}
(\byear{2015}).
doi:\doiurl{10.1093/mnras/stv226}
\end{barticle}
\endbibitem

\bibitem[\protect\citeauthoryear{{Budnik} et~al.}{2010}]{Budnik+10radshock}
\begin{barticle}
\bauthor{\binits{R.} \bsnm{{Budnik}}},
\bauthor{\binits{B.} \bsnm{{Katz}}},
\bauthor{\binits{A.} \bsnm{{Sagiv}}},
\bauthor{\binits{E.} \bsnm{{Waxman}}},
\batitle{{Relativistic Radiation Mediated Shocks}}.
\bjtitle{\apj}
\bvolume{725},
\bfpage{63}--\blpage{90}
(\byear{2010}).
doi:\doiurl{10.1088/0004-637X/725/1/63}
\end{barticle}
\endbibitem

\bibitem[\protect\citeauthoryear{{Burgess} et~al.}{2011}]{Burgess+11issy}
\begin{barticle}
\bauthor{\binits{J.M.} \bsnm{{Burgess}}},
\bauthor{\binits{R.D.} \bsnm{{Preece}}},
\bauthor{\binits{M.G.} \bsnm{{Baring}}},
\bauthor{\binits{M.S.} \bsnm{{Briggs}}},
\bauthor{\binits{V.} \bsnm{{Connaughton}}},
\bauthor{\binits{S.} \bsnm{{Guiriec}}},
\bauthor{\binits{W.S.} \bsnm{{Paciesas}}},
\bauthor{\binits{C.A.} \bsnm{{Meegan}}},
\bauthor{\binits{P.N.} \bsnm{{Bhat}}},
\bauthor{\binits{E.} \bsnm{{Bissaldi}}},
\bauthor{\binits{V.} \bsnm{{Chaplin}}},
\bauthor{\binits{R.} \bsnm{{Diehl}}},
\bauthor{\binits{G.J.} \bsnm{{Fishman}}},
\bauthor{\binits{G.} \bsnm{{Fitzpatrick}}},
\bauthor{\binits{S.} \bsnm{{Foley}}},
\bauthor{\binits{M.} \bsnm{{Gibby}}},
\bauthor{\binits{M.} \bsnm{{Giles}}},
\bauthor{\binits{A.} \bsnm{{Goldstein}}},
\bauthor{\binits{J.} \bsnm{{Greiner}}},
\bauthor{\binits{D.} \bsnm{{Gruber}}},
\bauthor{\binits{A.J.} \bsnm{{van der Horst}}},
\bauthor{\binits{A.} \bsnm{{von Kienlin}}},
\bauthor{\binits{M.} \bsnm{{Kippen}}},
\bauthor{\binits{C.} \bsnm{{Kouveliotou}}},
\bauthor{\binits{S.} \bsnm{{McBreen}}},
\bauthor{\binits{A.} \bsnm{{Rau}}},
\bauthor{\binits{D.} \bsnm{{Tierney}}},
\bauthor{\binits{C.} \bsnm{{Wilson-Hodge}}},
\batitle{{Constraints on the Synchrotron Shock Model for the Fermi GRB 090820A
  Observed by Gamma-Ray Burst Monitor}}.
\bjtitle{\apj}
\bvolume{741},
\bfpage{24}
(\byear{2011}).
doi:\doiurl{10.1088/0004-637X/741/1/24}
\end{barticle}
\endbibitem

\bibitem[\protect\citeauthoryear{{Burgess} et~al.}{2014}]{Burgess+14therm}
\begin{barticle}
\bauthor{\binits{J.M.} \bsnm{{Burgess}}},
\bauthor{\binits{R.D.} \bsnm{{Preece}}},
\bauthor{\binits{F.} \bsnm{{Ryde}}},
\bauthor{\binits{P.} \bsnm{{Veres}}},
\bauthor{\binits{P.} \bsnm{{M{\'e}sz{\'a}ros}}},
\bauthor{\binits{V.} \bsnm{{Connaughton}}},
\bauthor{\binits{M.} \bsnm{{Briggs}}},
\bauthor{\binits{A.} \bsnm{{Pe'er}}},
\bauthor{\binits{S.} \bsnm{{Iyyani}}},
\bauthor{\binits{A.} \bsnm{{Goldstein}}},
\bauthor{\binits{M.} \bsnm{{Axelsson}}},
\bauthor{\binits{M.G.} \bsnm{{Baring}}},
\bauthor{\binits{P.N.} \bsnm{{Bhat}}},
\bauthor{\binits{D.} \bsnm{{Byrne}}},
\bauthor{\binits{G.} \bsnm{{Fitzpatrick}}},
\bauthor{\binits{S.} \bsnm{{Foley}}},
\bauthor{\binits{D.} \bsnm{{Kocevski}}},
\bauthor{\binits{N.} \bsnm{{Omodei}}},
\bauthor{\binits{W.S.} \bsnm{{Paciesas}}},
\bauthor{\binits{V.} \bsnm{{Pelassa}}},
\bauthor{\binits{C.} \bsnm{{Kouveliotou}}},
\bauthor{\binits{S.} \bsnm{{Xiong}}},
\bauthor{\binits{H.-F.} \bsnm{{Yu}}},
\bauthor{\binits{B.} \bsnm{{Zhang}}},
\bauthor{\binits{S.} \bsnm{{Zhu}}},
\batitle{{An Observed Correlation between Thermal and Non-thermal Emission in
  Gamma-Ray Bursts}}.
\bjtitle{\apjl}
\bvolume{784},
\bfpage{43}
(\byear{2014}).
doi:\doiurl{10.1088/2041-8205/784/2/L43}
\end{barticle}
\endbibitem

\bibitem[\protect\citeauthoryear{{Daigne} and
  {Mochkovitch}}{1998}]{Daigne+98is}
\begin{barticle}
\bauthor{\binits{F.} \bsnm{{Daigne}}},
\bauthor{\binits{R.} \bsnm{{Mochkovitch}}},
\batitle{{Gamma-ray bursts from internal shocks in a relativistic wind:
  temporal and spectral properties}}.
\bjtitle{\mnras}
\bvolume{296},
\bfpage{275}--\blpage{286}
(\byear{1998}).
doi:\doiurl{10.1046/j.1365-8711.1998.01305.x}
\end{barticle}
\endbibitem

\bibitem[\protect\citeauthoryear{{Derishev} et~al.}{1999}]{Derishev+99}
\begin{barticle}
\bauthor{\binits{E.V.} \bsnm{{Derishev}}},
\bauthor{\binits{V.V.} \bsnm{{Kocharovsky}}},
\bauthor{\binits{V.V.} \bsnm{{Kocharovsky}}},
\batitle{{The Neutron Component in Fireballs of Gamma-Ray Bursts: Dynamics and
  Observable Imprints}}.
\bjtitle{\apj}
\bvolume{521},
\bfpage{640}--\blpage{649}
(\byear{1999}).
doi:\doiurl{10.1086/307574}
\end{barticle}
\endbibitem

\bibitem[\protect\citeauthoryear{{Derishev}
  et~al.}{2003}]{Derishev+03converter}
\begin{barticle}
\bauthor{\binits{E.V.} \bsnm{{Derishev}}},
\bauthor{\binits{F.A.} \bsnm{{Aharonian}}},
\bauthor{\binits{V.V.} \bsnm{{Kocharovsky}}},
\bauthor{\binits{V.V.} \bsnm{{Kocharovsky}}},
\batitle{{Particle acceleration through multiple conversions from a charged
  into a neutral state and back}}.
\bjtitle{\prd}
\bvolume{68}(\bissue{4}),
\bfpage{043003}
(\byear{2003}).
doi:\doiurl{10.1103/PhysRevD.68.043003}
\end{barticle}
\endbibitem

\bibitem[\protect\citeauthoryear{{Drenkhahn} and {Spruit}}{2002}]{Drenkhahn+02}
\begin{barticle}
\bauthor{\binits{G.} \bsnm{{Drenkhahn}}},
\bauthor{\binits{H.C.} \bsnm{{Spruit}}},
\batitle{{Efficient acceleration and radiation in Poynting flux powered GRB
  outflows}}.
\bjtitle{\aap}
\bvolume{391},
\bfpage{1141}--\blpage{1153}
(\byear{2002}).
doi:\doiurl{10.1051/0004-6361:20020839}
\end{barticle}
\endbibitem

\bibitem[\protect\citeauthoryear{{Eichler} and
  {Levinson}}{2000}]{Eichler+00thermal}
\begin{barticle}
\bauthor{\binits{D.} \bsnm{{Eichler}}},
\bauthor{\binits{A.} \bsnm{{Levinson}}},
\batitle{{A Compact Fireball Model of Gamma-Ray Bursts}}.
\bjtitle{\apj}
\bvolume{529},
\bfpage{146}--\blpage{150}
(\byear{2000}).
doi:\doiurl{10.1086/308245}
\end{barticle}
\endbibitem

\bibitem[\protect\citeauthoryear{{Ghirlanda}
  et~al.}{2012}]{Ghirlanda+12-Gamma-eiso}
\begin{barticle}
\bauthor{\binits{G.} \bsnm{{Ghirlanda}}},
\bauthor{\binits{L.} \bsnm{{Nava}}},
\bauthor{\binits{G.} \bsnm{{Ghisellini}}},
\bauthor{\binits{A.} \bsnm{{Celotti}}},
\bauthor{\binits{D.} \bsnm{{Burlon}}},
\bauthor{\binits{S.} \bsnm{{Covino}}},
\bauthor{\binits{A.} \bsnm{{Melandri}}},
\batitle{{Gamma-ray bursts in the comoving frame}}.
\bjtitle{\mnras}
\bvolume{420},
\bfpage{483}--\blpage{494}
(\byear{2012}).
doi:\doiurl{10.1111/j.1365-2966.2011.20053.x}
\end{barticle}
\endbibitem

\bibitem[\protect\citeauthoryear{{Ghisellini}}{2006}]{Ghisellini+06blazgrb}
\begin{bchapter}
\bauthor{\binits{G.} \bsnm{{Ghisellini}}},
\bctitle{{Blazars and Gamma Ray Bursts}},
in \bbtitle{VI Microquasar Workshop: Microquasars and Beyond},
\byear{2006},
pp. \bfpage{27}--\blpage{1}
\end{bchapter}
\endbibitem

\bibitem[\protect\citeauthoryear{{Giacinti} and {Bell}}{2015}]{Giacinti+15nusn}
\begin{barticle}
\bauthor{\binits{G.} \bsnm{{Giacinti}}},
\bauthor{\binits{A.R.} \bsnm{{Bell}}},
\batitle{{Collisionless shocks and TeV neutrinos before Supernova shock
  breakout from an optically thick wind}}.
\bjtitle{\mnras}
\bvolume{449},
\bfpage{3693}--\blpage{3699}
(\byear{2015}).
doi:\doiurl{10.1093/mnras/stv561}
\end{barticle}
\endbibitem

\bibitem[\protect\citeauthoryear{{Giannios}}{2008}]{Giannios08phot}
\begin{barticle}
\bauthor{\binits{D.} \bsnm{{Giannios}}},
\batitle{{Prompt GRB emission from gradual energy dissipation}}.
\bjtitle{\aap}
\bvolume{480},
\bfpage{305}--\blpage{312}
(\byear{2008}).
doi:\doiurl{10.1051/0004-6361:20079085}
\end{barticle}
\endbibitem

\bibitem[\protect\citeauthoryear{{Giannios}}{2012}]{Giannios12peak}
\begin{barticle}
\bauthor{\binits{D.} \bsnm{{Giannios}}},
\batitle{{The peak energy of dissipative gamma-ray burst photospheres}}.
\bjtitle{\mnras}
\bvolume{422},
\bfpage{3092}--\blpage{3098}
(\byear{2012}).
doi:\doiurl{10.1111/j.1365-2966.2012.20825.x}
\end{barticle}
\endbibitem

\bibitem[\protect\citeauthoryear{{Giannios} and
  {Spruit}}{2007}]{Giannios+07photspec}
\begin{barticle}
\bauthor{\binits{D.} \bsnm{{Giannios}}},
\bauthor{\binits{H.C.} \bsnm{{Spruit}}},
\batitle{{Spectral and timing properties of a dissipative gamma-ray burst
  photosphere}}.
\bjtitle{\aap}
\bvolume{469},
\bfpage{1}--\blpage{9}
(\byear{2007}).
doi:\doiurl{10.1051/0004-6361:20066739}
\end{barticle}
\endbibitem

\bibitem[\protect\citeauthoryear{{Gill} and {Thompson}}{2014}]{Gill+14pairphot}
\begin{barticle}
\bauthor{\binits{R.} \bsnm{{Gill}}},
\bauthor{\binits{C.} \bsnm{{Thompson}}},
\batitle{{Non-thermal Gamma-Ray Emission from Delayed Pair Breakdown in a
  Magnetized and Photon-rich Outflow}}.
\bjtitle{\apj}
\bvolume{796},
\bfpage{81}
(\byear{2014}).
doi:\doiurl{10.1088/0004-637X/796/2/81}
\end{barticle}
\endbibitem

\bibitem[\protect\citeauthoryear{{Goldstein}
  et~al.}{2012}]{Goldstein+12fermispcat}
\begin{barticle}
\bauthor{\binits{A.} \bsnm{{Goldstein}}},
\bauthor{\binits{J.M.} \bsnm{{Burgess}}},
\bauthor{\binits{R.D.} \bsnm{{Preece}}},
\bauthor{\binits{M.S.} \bsnm{{Briggs}}},
\bauthor{\binits{S.} \bsnm{{Guiriec}}},
\bauthor{\binits{A.J.} \bsnm{{van der Horst}}},
\bauthor{\binits{V.} \bsnm{{Connaughton}}},
\bauthor{\binits{C.A.} \bsnm{{Wilson-Hodge}}},
\bauthor{\binits{W.S.} \bsnm{{Paciesas}}},
\bauthor{\binits{C.A.} \bsnm{{Meegan}}},
\bauthor{\binits{A.} \bsnm{{von Kienlin}}},
\bauthor{\binits{P.N.} \bsnm{{Bhat}}},
\bauthor{\binits{E.} \bsnm{{Bissaldi}}},
\bauthor{\binits{V.} \bsnm{{Chaplin}}},
\bauthor{\binits{R.} \bsnm{{Diehl}}},
\bauthor{\binits{G.J.} \bsnm{{Fishman}}},
\bauthor{\binits{G.} \bsnm{{Fitzpatrick}}},
\bauthor{\binits{S.} \bsnm{{Foley}}},
\bauthor{\binits{M.} \bsnm{{Gibby}}},
\bauthor{\binits{M.} \bsnm{{Giles}}},
\bauthor{\binits{J.} \bsnm{{Greiner}}},
\bauthor{\binits{D.} \bsnm{{Gruber}}},
\bauthor{\binits{R.M.} \bsnm{{Kippen}}},
\bauthor{\binits{C.} \bsnm{{Kouveliotou}}},
\bauthor{\binits{S.} \bsnm{{McBreen}}},
\bauthor{\binits{S.} \bsnm{{McGlynn}}},
\bauthor{\binits{A.} \bsnm{{Rau}}},
\bauthor{\binits{D.} \bsnm{{Tierney}}},
\batitle{{The Fermi GBM Gamma-Ray Burst Spectral Catalog: The First Two
  Years}}.
\bjtitle{\apjs}
\bvolume{199},
\bfpage{19}
(\byear{2012}).
doi:\doiurl{10.1088/0067-0049/199/1/19}
\end{barticle}
\endbibitem

\bibitem[\protect\citeauthoryear{{Goodman}}{1986}]{Goodman86grb}
\begin{barticle}
\bauthor{\binits{J.} \bsnm{{Goodman}}},
\batitle{{Are gamma-ray bursts optically thick?}}
\bjtitle{\apjl}
\bvolume{308},
\bfpage{47}--\blpage{50}
(\byear{1986}).
doi:\doiurl{10.1086/184741}
\end{barticle}
\endbibitem

\bibitem[\protect\citeauthoryear{{G{\"o}tz} et~al.}{2009}]{GotEtAl:2009}
\begin{barticle}
\bauthor{\binits{D.} \bsnm{{G{\"o}tz}}},
\bauthor{\binits{P.} \bsnm{{Laurent}}},
\bauthor{\binits{F.} \bsnm{{Lebrun}}},
\bauthor{\binits{F.} \bsnm{{Daigne}}},
\bauthor{\binits{{\v Z}.} \bsnm{{Bo{\v s}njak}}},
\batitle{{Variable Polarization Measured in the Prompt Emission of GRB 041219A
  Using IBIS on Board INTEGRAL}}.
\bjtitle{\apjl}
\bvolume{695},
\bfpage{208}--\blpage{212}
(\byear{2009}).
doi:\doiurl{10.1088/0004-637X/695/2/L208}
\end{barticle}
\endbibitem

\bibitem[\protect\citeauthoryear{{G{\"o}tz} et~al.}{2013}]{GotEtAl:2013}
\begin{barticle}
\bauthor{\binits{D.} \bsnm{{G{\"o}tz}}},
\bauthor{\binits{S.} \bsnm{{Covino}}},
\bauthor{\binits{A.} \bsnm{{Fern{\'a}ndez-Soto}}},
\bauthor{\binits{P.} \bsnm{{Laurent}}},
\bauthor{\binits{{\v Z}.} \bsnm{{Bo{\v s}njak}}},
\batitle{{The polarized gamma-ray burst GRB 061122}}.
\bjtitle{\mnras}
\bvolume{431},
\bfpage{3550}--\blpage{3556}
(\byear{2013}).
doi:\doiurl{10.1093/mnras/stt439}
\end{barticle}
\endbibitem

\bibitem[\protect\citeauthoryear{{G{\"o}tz} et~al.}{2014}]{GotEtAl:2014}
\begin{barticle}
\bauthor{\binits{D.} \bsnm{{G{\"o}tz}}},
\bauthor{\binits{P.} \bsnm{{Laurent}}},
\bauthor{\binits{S.} \bsnm{{Antier}}},
\bauthor{\binits{S.} \bsnm{{Covino}}},
\bauthor{\binits{P.} \bsnm{{D'Avanzo}}},
\bauthor{\binits{V.} \bsnm{{D'Elia}}},
\bauthor{\binits{A.} \bsnm{{Melandri}}},
\batitle{{GRB 140206A: the most distant polarized gamma-ray burst}}.
\bjtitle{\mnras}
\bvolume{444},
\bfpage{2776}--\blpage{2782}
(\byear{2014}).
doi:\doiurl{10.1093/mnras/stu1634}
\end{barticle}
\endbibitem

\bibitem[\protect\citeauthoryear{{Hasco{\"e}t} et~al.}{2015}]{Hascoet+15grblf}
\begin{barticle}
\bauthor{\binits{R.} \bsnm{{Hasco{\"e}t}}},
\bauthor{\binits{I.} \bsnm{{Vurm}}},
\bauthor{\binits{A.M.} \bsnm{{Beloborodov}}},
\batitle{{Measuring Ambient Densities and Lorentz Factors of Gamma-Ray Bursts
  from GeV and Optical Observations}}.
\bjtitle{\apj}
\bvolume{813},
\bfpage{63}
(\byear{2015}).
doi:\doiurl{10.1088/0004-637X/813/1/63}
\end{barticle}
\endbibitem

\bibitem[\protect\citeauthoryear{{Ito} et~al.}{2014}]{ItoEtAl:2014}
\begin{barticle}
\bauthor{\binits{H.} \bsnm{{Ito}}},
\bauthor{\binits{S.} \bsnm{{Nagataki}}},
\bauthor{\binits{J.} \bsnm{{Matsumoto}}},
\bauthor{\binits{S.-H.} \bsnm{{Lee}}},
\bauthor{\binits{A.} \bsnm{{Tolstov}}},
\bauthor{\binits{J.} \bsnm{{Mao}}},
\bauthor{\binits{M.} \bsnm{{Dainotti}}},
\bauthor{\binits{A.} \bsnm{{Mizuta}}},
\batitle{{Spectral and Polarization Properties of Photospheric Emission from
  Stratified Jets}}.
\bjtitle{\apj}
\bvolume{789},
\bfpage{159}
(\byear{2014}).
doi:\doiurl{10.1088/0004-637X/789/2/159}
\end{barticle}
\endbibitem

\bibitem[\protect\citeauthoryear{{Ito} et~al.}{2015}]{Ito+15photjet}
\begin{barticle}
\bauthor{\binits{H.} \bsnm{{Ito}}},
\bauthor{\binits{J.} \bsnm{{Matsumoto}}},
\bauthor{\binits{S.} \bsnm{{Nagataki}}},
\bauthor{\binits{D.C.} \bsnm{{Warren}}},
\bauthor{\binits{M.V.} \bsnm{{Barkov}}},
\batitle{{Photospheric Emission from Collapsar Jets in 3D Relativistic
  Hydrodynamics}}.
\bjtitle{\apjl}
\bvolume{814},
\bfpage{29}
(\byear{2015}).
doi:\doiurl{10.1088/2041-8205/814/2/L29}
\end{barticle}
\endbibitem

\bibitem[\protect\citeauthoryear{{Kagan} et~al.}{2015}]{Kagan+15recon}
\begin{barticle}
\bauthor{\binits{D.} \bsnm{{Kagan}}},
\bauthor{\binits{L.} \bsnm{{Sironi}}},
\bauthor{\binits{B.} \bsnm{{Cerutti}}},
\bauthor{\binits{D.} \bsnm{{Giannios}}},
\batitle{{Relativistic Magnetic Reconnection in Pair Plasmas and Its
  Astrophysical Applications}}.
\bjtitle{SSRv.}
\bvolume{191},
\bfpage{545}--\blpage{573}
(\byear{2015}).
doi:\doiurl{10.1007/s11214-014-0132-9}
\end{barticle}
\endbibitem

\bibitem[\protect\citeauthoryear{{Kalemci} et~al.}{2007}]{KalEtAl:2007}
\begin{barticle}
\bauthor{\binits{E.} \bsnm{{Kalemci}}},
\bauthor{\binits{S.E.} \bsnm{{Boggs}}},
\bauthor{\binits{C.} \bsnm{{Kouveliotou}}},
\bauthor{\binits{M.} \bsnm{{Finger}}},
\bauthor{\binits{M.G.} \bsnm{{Baring}}},
\batitle{{Search for Polarization from the Prompt Gamma-Ray Emission of GRB
  041219a with SPI on INTEGRAL}}.
\bjtitle{\apjs}
\bvolume{169},
\bfpage{75}--\blpage{82}
(\byear{2007}).
doi:\doiurl{10.1086/510676}
\end{barticle}
\endbibitem

\bibitem[\protect\citeauthoryear{{Kaneko} et~al.}{2006}]{Kaneko+06batsespec}
\begin{barticle}
\bauthor{\binits{Y.} \bsnm{{Kaneko}}},
\bauthor{\binits{R.D.} \bsnm{{Preece}}},
\bauthor{\binits{M.S.} \bsnm{{Briggs}}},
\bauthor{\binits{W.S.} \bsnm{{Paciesas}}},
\bauthor{\binits{C.A.} \bsnm{{Meegan}}},
\bauthor{\binits{D.L.} \bsnm{{Band}}},
\batitle{{The Complete Spectral Catalog of Bright BATSE Gamma-Ray Bursts}}.
\bjtitle{\apjs}
\bvolume{166},
\bfpage{298}--\blpage{340}
(\byear{2006}).
doi:\doiurl{10.1086/505911}
\end{barticle}
\endbibitem

\bibitem[\protect\citeauthoryear{{Kashiyama} et~al.}{2013}]{Kashiyama+13pnconv}
\begin{barticle}
\bauthor{\binits{K.} \bsnm{{Kashiyama}}},
\bauthor{\binits{K.} \bsnm{{Murase}}},
\bauthor{\binits{P.} \bsnm{{M{\'e}sz{\'a}ros}}},
\batitle{{Neutron-Proton-Converter Acceleration Mechanism at Subphotospheres of
  Relativistic Outflows}}.
\bjtitle{Physical Review Letters}
\bvolume{111}(\bissue{13}),
\bfpage{131103}
(\byear{2013}).
doi:\doiurl{10.1103/PhysRevLett.111.131103}
\end{barticle}
\endbibitem

\bibitem[\protect\citeauthoryear{{Kobayashi} et~al.}{1997}]{Kobayashi+97variab}
\begin{barticle}
\bauthor{\binits{S.} \bsnm{{Kobayashi}}},
\bauthor{\binits{T.} \bsnm{{Piran}}},
\bauthor{\binits{R.} \bsnm{{Sari}}},
\batitle{{Can Internal Shocks Produce the Variability in Gamma-Ray Bursts?}}
\bjtitle{\apj}
\bvolume{490},
\bfpage{92}
(\byear{1997})
\end{barticle}
\endbibitem

\bibitem[\protect\citeauthoryear{{Komissarov}
  et~al.}{2009}]{Komissarov+09maggrb}
\begin{barticle}
\bauthor{\binits{S.S.} \bsnm{{Komissarov}}},
\bauthor{\binits{N.} \bsnm{{Vlahakis}}},
\bauthor{\binits{A.} \bsnm{{K{\"o}nigl}}},
\bauthor{\binits{M.V.} \bsnm{{Barkov}}},
\batitle{{Magnetic acceleration of ultrarelativistic jets in gamma-ray burst
  sources}}.
\bjtitle{\mnras}
\bvolume{394},
\bfpage{1182}--\blpage{1212}
(\byear{2009}).
doi:\doiurl{10.1111/j.1365-2966.2009.14410.x}
\end{barticle}
\endbibitem

\bibitem[\protect\citeauthoryear{{Lazzati} et~al.}{2009}]{Lazzati+09phot}
\begin{barticle}
\bauthor{\binits{D.} \bsnm{{Lazzati}}},
\bauthor{\binits{B.J.} \bsnm{{Morsony}}},
\bauthor{\binits{M.C.} \bsnm{{Begelman}}},
\batitle{{Very High Efficiency Photospheric Emission in Long-Duration
  {$\gamma$}-Ray Bursts}}.
\bjtitle{\apjl}
\bvolume{700},
\bfpage{47}--\blpage{50}
(\byear{2009}).
doi:\doiurl{10.1088/0004-637X/700/1/L47}
\end{barticle}
\endbibitem

\bibitem[\protect\citeauthoryear{{Lazzati} et~al.}{2013}]{Lazzati+13dur}
\begin{botherref}
\oauthor{\binits{D.} \bsnm{{Lazzati}}},
\oauthor{\binits{M.} \bsnm{{Villeneuve}}},
\oauthor{\binits{D.} \bsnm{{Lopez-Camara}}},
\oauthor{\binits{B.} \bsnm{{Morsony}}},
\oauthor{\binits{R.} \bsnm{{Perna}}},
{On the observed duration distribution of gamma-ray bursts from collapsars}.
ArXiv e-prints
(2013)
\end{botherref}
\endbibitem

\bibitem[\protect\citeauthoryear{{Levinson}}{2012}]{Levinson12radshock}
\begin{barticle}
\bauthor{\binits{A.} \bsnm{{Levinson}}},
\batitle{{Observational Signatures of Sub-photospheric Radiation-mediated
  Shocks in the Prompt Phase of Gamma-Ray Bursts}}.
\bjtitle{\apj}
\bvolume{756},
\bfpage{174}
(\byear{2012}).
doi:\doiurl{10.1088/0004-637X/756/2/174}
\end{barticle}
\endbibitem

\bibitem[\protect\citeauthoryear{{Levinson} and
  {Bromberg}}{2008}]{2008PhRvL.100m1101L}
\begin{barticle}
\bauthor{\binits{A.} \bsnm{{Levinson}}},
\bauthor{\binits{O.} \bsnm{{Bromberg}}},
\batitle{{Relativistic Photon Mediated Shocks}}.
\bjtitle{Physical Review Letters}
\bvolume{100}(\bissue{13}),
\bfpage{131101}
(\byear{2008}).
doi:\doiurl{10.1103/PhysRevLett.100.131101}
\end{barticle}
\endbibitem

\bibitem[\protect\citeauthoryear{{Lundman} et~al.}{2014}]{Lundman+14grbpol}
\begin{barticle}
\bauthor{\binits{C.} \bsnm{{Lundman}}},
\bauthor{\binits{A.} \bsnm{{Pe'er}}},
\bauthor{\binits{F.} \bsnm{{Ryde}}},
\batitle{{Polarization properties of photospheric emission from relativistic,
  collimated outflows}}.
\bjtitle{\mnras}
\bvolume{440},
\bfpage{3292}--\blpage{3308}
(\byear{2014}).
doi:\doiurl{10.1093/mnras/stu457}
\end{barticle}
\endbibitem

\bibitem[\protect\citeauthoryear{{Lundman} et~al.}{2016}]{Lundman+16}
\begin{botherref}
\oauthor{\binits{C.} \bsnm{{Lundman}}},
\oauthor{\binits{I.} \bsnm{{Vurm}}},
\oauthor{\binits{A.M.} \bsnm{{Beloborodov}}},
{Polarization of gamma-ray bursts in the dissipative photosphere model}.
ArXiv e-prints
(2016)
\end{botherref}
\endbibitem

\bibitem[\protect\citeauthoryear{{Lyutikov} et~al.}{2003}]{LyuParBla:2003}
\begin{barticle}
\bauthor{\binits{M.} \bsnm{{Lyutikov}}},
\bauthor{\binits{V.I.} \bsnm{{Pariev}}},
\bauthor{\binits{R.D.} \bsnm{{Blandford}}},
\batitle{{Polarization of Prompt Gamma-Ray Burst Emission: Evidence for
  Electromagnetically Dominated Outflow}}.
\bjtitle{\apj}
\bvolume{597},
\bfpage{998}--\blpage{1009}
(\byear{2003}).
doi:\doiurl{10.1086/378497}
\end{barticle}
\endbibitem

\bibitem[\protect\citeauthoryear{{McGlynn} et~al.}{2007}]{McGEtAl:2007}
\begin{barticle}
\bauthor{\binits{S.} \bsnm{{McGlynn}}},
\bauthor{\binits{D.J.} \bsnm{{Clark}}},
\bauthor{\binits{A.J.} \bsnm{{Dean}}},
\bauthor{\binits{L.} \bsnm{{Hanlon}}},
\bauthor{\binits{S.} \bsnm{{McBreen}}},
\bauthor{\binits{D.R.} \bsnm{{Willis}}},
\bauthor{\binits{B.} \bsnm{{McBreen}}},
\bauthor{\binits{A.J.} \bsnm{{Bird}}},
\bauthor{\binits{S.} \bsnm{{Foley}}},
\batitle{{Polarisation studies of the prompt gamma-ray emission from GRB
  041219a using the spectrometer aboard INTEGRAL}}.
\bjtitle{\aap}
\bvolume{466},
\bfpage{895}--\blpage{904}
(\byear{2007}).
doi:\doiurl{10.1051/0004-6361:20066179}
\end{barticle}
\endbibitem

\bibitem[\protect\citeauthoryear{{McGlynn} et~al.}{2009}]{McGEtAl:2009}
\begin{barticle}
\bauthor{\binits{S.} \bsnm{{McGlynn}}},
\bauthor{\binits{S.} \bsnm{{Foley}}},
\bauthor{\binits{B.} \bsnm{{McBreen}}},
\bauthor{\binits{L.} \bsnm{{Hanlon}}},
\bauthor{\binits{S.} \bsnm{{McBreen}}},
\bauthor{\binits{D.J.} \bsnm{{Clark}}},
\bauthor{\binits{A.J.} \bsnm{{Dean}}},
\bauthor{\binits{A.} \bsnm{{Martin-Carrillo}}},
\bauthor{\binits{R.} \bsnm{{O'Connor}}},
\batitle{{High energy emission and polarisation limits for the INTEGRAL burst
  GRB 061122}}.
\bjtitle{\aap}
\bvolume{499},
\bfpage{465}--\blpage{472}
(\byear{2009}).
doi:\doiurl{10.1051/0004-6361/200810920}
\end{barticle}
\endbibitem

\bibitem[\protect\citeauthoryear{{M{\'e}sz{\'a}ros} and
  {Rees}}{2000a}]{Meszaros+00gevnu}
\begin{barticle}
\bauthor{\binits{P.} \bsnm{{M{\'e}sz{\'a}ros}}},
\bauthor{\binits{M.J.} \bsnm{{Rees}}},
\batitle{{Multi-GEV Neutrinos from Internal Dissipation in Gamma-Ray Burst
  Fireballs}}.
\bjtitle{\apjl}
\bvolume{541},
\bfpage{5}--\blpage{8}
(\byear{2000}a).
doi:\doiurl{10.1086/312894}
\end{barticle}
\endbibitem

\bibitem[\protect\citeauthoryear{{M{\'e}sz{\'a}ros} and
  {Rees}}{2000b}]{Meszaros+00phot}
\begin{barticle}
\bauthor{\binits{P.} \bsnm{{M{\'e}sz{\'a}ros}}},
\bauthor{\binits{M.J.} \bsnm{{Rees}}},
\batitle{{Steep Slopes and Preferred Breaks in Gamma-Ray Burst Spectra: The
  Role of Photospheres and Comptonization}}.
\bjtitle{\apj}
\bvolume{530},
\bfpage{292}--\blpage{298}
(\byear{2000}b).
doi:\doiurl{10.1086/308371}
\end{barticle}
\endbibitem

\bibitem[\protect\citeauthoryear{{M{\'e}sz{\'a}ros} and
  {Waxman}}{2001}]{Meszaros+01choked}
\begin{barticle}
\bauthor{\binits{P.} \bsnm{{M{\'e}sz{\'a}ros}}},
\bauthor{\binits{E.} \bsnm{{Waxman}}},
\batitle{{TeV Neutrinos from Successful and Choked Gamma-Ray Bursts}}.
\bjtitle{Physical Review Letters}
\bvolume{87}(\bissue{17}),
\bfpage{171102}
(\byear{2001})
\end{barticle}
\endbibitem

\bibitem[\protect\citeauthoryear{{Morsony} et~al.}{2010}]{Morsony+10variab}
\begin{barticle}
\bauthor{\binits{B.J.} \bsnm{{Morsony}}},
\bauthor{\binits{D.} \bsnm{{Lazzati}}},
\bauthor{\binits{M.C.} \bsnm{{Begelman}}},
\batitle{{The Origin and Propagation of Variability in the Outflows of
  Long-duration Gamma-ray Bursts}}.
\bjtitle{\apj}
\bvolume{723},
\bfpage{267}--\blpage{276}
(\byear{2010}).
doi:\doiurl{10.1088/0004-637X/723/1/267}
\end{barticle}
\endbibitem

\bibitem[\protect\citeauthoryear{{Murase} et~al.}{2013}]{Murase+13subphotnu}
\begin{barticle}
\bauthor{\binits{K.} \bsnm{{Murase}}},
\bauthor{\binits{K.} \bsnm{{Kashiyama}}},
\bauthor{\binits{P.} \bsnm{{M{\'e}sz{\'a}ros}}},
\batitle{{Subphotospheric Neutrinos from Gamma-Ray Bursts: The Role of
  Neutrons}}.
\bjtitle{Physical Review Letters}
\bvolume{111}(\bissue{13}),
\bfpage{131102}
(\byear{2013}).
doi:\doiurl{10.1103/PhysRevLett.111.131102}
\end{barticle}
\endbibitem

\bibitem[\protect\citeauthoryear{{Pacz\'ynski}}{1986}]{Paczynski86}
\begin{barticle}
\bauthor{\binits{B.} \bsnm{{Pacz\'ynski}}},
\batitle{{Gamma-ray bursters at cosmological distances}}.
\bjtitle{\apjl}
\bvolume{308},
\bfpage{43}--\blpage{46}
(\byear{1986}).
doi:\doiurl{10.1086/184740}
\end{barticle}
\endbibitem

\bibitem[\protect\citeauthoryear{{Pe'er}}{2008}]{Peer_2008}
\begin{barticle}
\bauthor{\binits{A.} \bsnm{{Pe'er}}},
\batitle{{Temporal Evolution of Thermal Emission from Relativistically
  Expanding Plasma}}.
\bjtitle{\apj}
\bvolume{682},
\bfpage{463}--\blpage{473}
(\byear{2008}).
doi:\doiurl{10.1086/588136}
\end{barticle}
\endbibitem

\bibitem[\protect\citeauthoryear{{Pe'er}}{2016}]{Peer+16}
\begin{bchapter}
\bauthor{\binits{A.} \bsnm{{Pe'er}}},
\bctitle{{''Photospheric Emission in Gamma-Ray Bursts''}},
in \bbtitle{41st COSPAR Scientific Assembly}.
\bsertitle{COSPAR Meeting},
vol. \bseriesno{41},
\byear{2016}
\end{bchapter}
\endbibitem

\bibitem[\protect\citeauthoryear{{Pe'er} et~al.}{2006}]{Peer+06phot}
\begin{barticle}
\bauthor{\binits{A.} \bsnm{{Pe'er}}},
\bauthor{\binits{P.} \bsnm{{M{\'e}sz{\'a}ros}}},
\bauthor{\binits{M.J.} \bsnm{{Rees}}},
\batitle{{The Observable Effects of a Photospheric Component on GRB and XRF
  Prompt Emission Spectrum}}.
\bjtitle{\apj}
\bvolume{642},
\bfpage{995}--\blpage{1003}
(\byear{2006}).
doi:\doiurl{10.1086/501424}
\end{barticle}
\endbibitem

\bibitem[\protect\citeauthoryear{{Pe'Er} et~al.}{2012}]{Peer+12-090902b}
\begin{barticle}
\bauthor{\binits{A.} \bsnm{{Pe'Er}}},
\bauthor{\binits{B.-B.} \bsnm{{Zhang}}},
\bauthor{\binits{F.} \bsnm{{Ryde}}},
\bauthor{\binits{S.} \bsnm{{McGlynn}}},
\bauthor{\binits{B.} \bsnm{{Zhang}}},
\bauthor{\binits{R.D.} \bsnm{{Preece}}},
\bauthor{\binits{C.} \bsnm{{Kouveliotou}}},
\batitle{{The connection between thermal and non-thermal emission in gamma-ray
  bursts: general considerations and GRB 090902B as a case study}}.
\bjtitle{\mnras}
\bvolume{420},
\bfpage{468}--\blpage{482}
(\byear{2012}).
doi:\doiurl{10.1111/j.1365-2966.2011.20052.x}
\end{barticle}
\endbibitem

\bibitem[\protect\citeauthoryear{{Preece} and {the Fermi
  collab.}}{2014}]{Preece+14-130427}
\begin{barticle}
\bauthor{\binits{R.} \bsnm{{Preece}}},
\bauthor{\bsnm{{the Fermi collab.}}},
\batitle{{The First Pulse of the Extremely Bright GRB 130427A: A Test Lab for
  Synchrotron Shocks}}.
\bjtitle{Science}
\bvolume{343},
\bfpage{51}--\blpage{54}
(\byear{2014}).
doi:\doiurl{10.1126/science.1242302}
\end{barticle}
\endbibitem

\bibitem[\protect\citeauthoryear{Razzaque et~al.}{2003}]{Razzaque+03nutomo}
\begin{barticle}
\bauthor{\binits{S.} \bsnm{Razzaque}},
\bauthor{\binits{P.} \bsnm{{M{\'e}sz{\'a}ros}}},
\bauthor{\binits{E.} \bsnm{Waxman}},
\batitle{Neutrino tomography of gamma ray bursts and massive stellar
  collapses}.
\bjtitle{Phys. Rev.}
\bvolume{D68},
\bfpage{083001}
(\byear{2003})
\end{barticle}
\endbibitem

\bibitem[\protect\citeauthoryear{{Rees} and {M\'esz\'aros}}{1994}]{Rees+94is}
\begin{barticle}
\bauthor{\binits{M.J.} \bsnm{{Rees}}},
\bauthor{\binits{P.} \bsnm{{M\'esz\'aros}}},
\batitle{{Unsteady outflow models for cosmological gamma-ray bursts}}.
\bjtitle{\apjl}
\bvolume{430},
\bfpage{93}--\blpage{96}
(\byear{1994}).
doi:\doiurl{10.1086/187446}
\end{barticle}
\endbibitem

\bibitem[\protect\citeauthoryear{{Rees} and
  {M{\'e}sz{\'a}ros}}{2005}]{Rees+05photdis}
\begin{barticle}
\bauthor{\binits{M.J.} \bsnm{{Rees}}},
\bauthor{\binits{P.} \bsnm{{M{\'e}sz{\'a}ros}}},
\batitle{{Dissipative Photosphere Models of Gamma-Ray Bursts and X-Ray
  Flashes}}.
\bjtitle{\apj}
\bvolume{628},
\bfpage{847}--\blpage{852}
(\byear{2005}).
doi:\doiurl{10.1086/430818}
\end{barticle}
\endbibitem

\bibitem[\protect\citeauthoryear{{Riffert}}{1988}]{Riffert88radshock}
\begin{barticle}
\bauthor{\binits{H.} \bsnm{{Riffert}}},
\batitle{{A self-consistent shock solution for radiation-dominated flows}}.
\bjtitle{\apj}
\bvolume{327},
\bfpage{760}--\blpage{771}
(\byear{1988}).
doi:\doiurl{10.1086/166234}
\end{barticle}
\endbibitem

\bibitem[\protect\citeauthoryear{{Rossi} et~al.}{2006}]{2006MNRAS.369.1797R}
\begin{barticle}
\bauthor{\binits{E.M.} \bsnm{{Rossi}}},
\bauthor{\binits{A.M.} \bsnm{{Beloborodov}}},
\bauthor{\binits{M.J.} \bsnm{{Rees}}},
\batitle{{Neutron-loaded outflows in gamma-ray bursts}}.
\bjtitle{\mnras}
\bvolume{369},
\bfpage{1797}--\blpage{1807}
(\byear{2006}).
doi:\doiurl{10.1111/j.1365-2966.2006.10417.x}
\end{barticle}
\endbibitem

\bibitem[\protect\citeauthoryear{{Russo} and
  {Thompson}}{2013}]{Russo+13magjet1}
\begin{barticle}
\bauthor{\binits{M.} \bsnm{{Russo}}},
\bauthor{\binits{C.} \bsnm{{Thompson}}},
\batitle{{Hot Electromagnetic Outflows. I. Acceleration and Spectra}}.
\bjtitle{\apj}
\bvolume{767},
\bfpage{142}
(\byear{2013}).
doi:\doiurl{10.1088/0004-637X/767/2/142}
\end{barticle}
\endbibitem

\bibitem[\protect\citeauthoryear{{Rybicki} and {Lightman}}{1979}]{Rybicki_book}
\begin{bbook}
\bauthor{\binits{G.B.} \bsnm{{Rybicki}}},
\bauthor{\binits{A.P.} \bsnm{{Lightman}}},
\bbtitle{{Radiative processes in astrophysics}}
\byear{1979}
\end{bbook}
\endbibitem

\bibitem[\protect\citeauthoryear{{Ryde}}{2004}]{Ryde04}
\begin{barticle}
\bauthor{\binits{F.} \bsnm{{Ryde}}},
\batitle{{The Cooling Behavior of Thermal Pulses in Gamma-Ray Bursts}}.
\bjtitle{\apj}
\bvolume{614},
\bfpage{827}--\blpage{846}
(\byear{2004}).
doi:\doiurl{10.1086/423782}
\end{barticle}
\endbibitem

\bibitem[\protect\citeauthoryear{{Ryde} et~al.}{2011}]{Ryde+11phot}
\begin{barticle}
\bauthor{\binits{F.} \bsnm{{Ryde}}},
\bauthor{\binits{A.} \bsnm{{Pe'Er}}},
\bauthor{\binits{T.} \bsnm{{Nymark}}},
\bauthor{\binits{M.} \bsnm{{Axelsson}}},
\bauthor{\binits{E.} \bsnm{{Moretti}}},
\bauthor{\binits{C.} \bsnm{{Lundman}}},
\bauthor{\binits{M.} \bsnm{{Battelino}}},
\bauthor{\binits{E.} \bsnm{{Bissaldi}}},
\bauthor{\binits{J.} \bsnm{{Chiang}}},
\bauthor{\binits{M.S.} \bsnm{{Jackson}}},
\bauthor{\binits{S.} \bsnm{{Larsson}}},
\bauthor{\binits{F.} \bsnm{{Longo}}},
\bauthor{\binits{S.} \bsnm{{McGlynn}}},
\bauthor{\binits{N.} \bsnm{{Omodei}}},
\batitle{{Observational evidence of dissipative photospheres in gamma-ray
  bursts}}.
\bjtitle{\mnras}
\bvolume{415},
\bfpage{3693}--\blpage{3705}
(\byear{2011}).
doi:\doiurl{10.1111/j.1365-2966.2011.18985.x}
\end{barticle}
\endbibitem

\bibitem[\protect\citeauthoryear{{Sironi} and
  {Spitkovsky}}{2011}]{Sironi+11magshockei}
\begin{barticle}
\bauthor{\binits{L.} \bsnm{{Sironi}}},
\bauthor{\binits{A.} \bsnm{{Spitkovsky}}},
\batitle{{Particle Acceleration in Relativistic Magnetized Collisionless
  Electron-Ion Shocks}}.
\bjtitle{\apj}
\bvolume{726},
\bfpage{75}
(\byear{2011}).
doi:\doiurl{10.1088/0004-637X/726/2/75}
\end{barticle}
\endbibitem

\bibitem[\protect\citeauthoryear{{Thompson}}{1994}]{Thompson94}
\begin{barticle}
\bauthor{\binits{C.} \bsnm{{Thompson}}},
\batitle{{A Model of Gamma-Ray Bursts}}.
\bjtitle{\mnras}
\bvolume{270},
\bfpage{480}
(\byear{1994})
\end{barticle}
\endbibitem

\bibitem[\protect\citeauthoryear{{Thompson} and {Gill}}{2014}]{Thompson+14phot}
\begin{barticle}
\bauthor{\binits{C.} \bsnm{{Thompson}}},
\bauthor{\binits{R.} \bsnm{{Gill}}},
\batitle{{Hot Electromagnetic Outflows. III. Displaced Fireball in a Strong
  Magnetic Field}}.
\bjtitle{\apj}
\bvolume{791},
\bfpage{46}
(\byear{2014}).
doi:\doiurl{10.1088/0004-637X/791/1/46}
\end{barticle}
\endbibitem

\bibitem[\protect\citeauthoryear{{Thompson} et~al.}{2007}]{Thompson+07phot}
\begin{barticle}
\bauthor{\binits{C.} \bsnm{{Thompson}}},
\bauthor{\binits{P.} \bsnm{{M{\'e}sz{\'a}ros}}},
\bauthor{\binits{M.J.} \bsnm{{Rees}}},
\batitle{{Thermalization in Relativistic Outflows and the Correlation between
  Spectral Hardness and Apparent Luminosity in Gamma-Ray Bursts}}.
\bjtitle{\apj}
\bvolume{666},
\bfpage{1012}--\blpage{1023}
(\byear{2007}).
doi:\doiurl{10.1086/518551}
\end{barticle}
\endbibitem

\bibitem[\protect\citeauthoryear{{Vlahakis} and
  {K{\"o}nigl}}{2003}]{Vlahakis+03grbmhd1}
\begin{barticle}
\bauthor{\binits{N.} \bsnm{{Vlahakis}}},
\bauthor{\binits{A.} \bsnm{{K{\"o}nigl}}},
\batitle{{Relativistic Magnetohydrodynamics with Application to Gamma-Ray Burst
  Outflows. I. Theory and Semianalytic Trans-Alfv{\'e}nic Solutions}}.
\bjtitle{\apj}
\bvolume{596},
\bfpage{1080}--\blpage{1103}
(\byear{2003}).
doi:\doiurl{10.1086/378226}
\end{barticle}
\endbibitem

\bibitem[\protect\citeauthoryear{{Vurm} and {Beloborodov}}{2016a}]{Vurm+TeV16}
\begin{botherref}
\oauthor{\binits{I.} \bsnm{{Vurm}}},
\oauthor{\binits{A.M.} \bsnm{{Beloborodov}}},
{On the prospects of gamma-ray burst detection in the TeV band}.
ArXiv e-prints
(2016a)
\end{botherref}
\endbibitem

\bibitem[\protect\citeauthoryear{{Vurm} and
  {Beloborodov}}{2016b}]{Vurm+16grbphot}
\begin{barticle}
\bauthor{\binits{I.} \bsnm{{Vurm}}},
\bauthor{\binits{A.M.} \bsnm{{Beloborodov}}},
\batitle{{Radiative Transfer Models for Gamma-Ray Bursts}}.
\bjtitle{\apj}
\bvolume{831},
\bfpage{175}
(\byear{2016}b).
doi:\doiurl{10.3847/0004-637X/831/2/175}
\end{barticle}
\endbibitem

\bibitem[\protect\citeauthoryear{{Vurm} et~al.}{2011}]{Vurm+11phot}
\begin{barticle}
\bauthor{\binits{I.} \bsnm{{Vurm}}},
\bauthor{\binits{A.M.} \bsnm{{Beloborodov}}},
\bauthor{\binits{J.} \bsnm{{Poutanen}}},
\batitle{{Gamma-Ray Bursts from Magnetized Collisionally Heated Jets}}.
\bjtitle{\apj}
\bvolume{738},
\bfpage{77}
(\byear{2011}).
doi:\doiurl{10.1088/0004-637X/738/1/77}
\end{barticle}
\endbibitem

\bibitem[\protect\citeauthoryear{{Vurm} et~al.}{2014}]{Vurm+14pairgevopt}
\begin{barticle}
\bauthor{\binits{I.} \bsnm{{Vurm}}},
\bauthor{\binits{R.} \bsnm{{Hasco{\"e}t}}},
\bauthor{\binits{A.M.} \bsnm{{Beloborodov}}},
\batitle{{Pair-dominated GeV-Optical Flash in GRB 130427A}}.
\bjtitle{\apjl}
\bvolume{789},
\bfpage{37}
(\byear{2014}).
doi:\doiurl{10.1088/2041-8205/789/2/L37}
\end{barticle}
\endbibitem

\bibitem[\protect\citeauthoryear{{Vurm} et~al.}{2013}]{Vurm+13phot}
\begin{barticle}
\bauthor{\binits{I.} \bsnm{{Vurm}}},
\bauthor{\binits{Y.} \bsnm{{Lyubarsky}}},
\bauthor{\binits{T.} \bsnm{{Piran}}},
\batitle{{On Thermalization in Gamma-Ray Burst Jets and the Peak Energies of
  Photospheric Spectra}}.
\bjtitle{\apj}
\bvolume{764},
\bfpage{143}
(\byear{2013}).
doi:\doiurl{10.1088/0004-637X/764/2/143}
\end{barticle}
\endbibitem

\bibitem[\protect\citeauthoryear{{Waxman} and {Loeb}}{2001}]{Waxman+01nusn}
\begin{barticle}
\bauthor{\binits{E.} \bsnm{{Waxman}}},
\bauthor{\binits{A.} \bsnm{{Loeb}}},
\batitle{{TeV Neutrinos and GeV Photons from Shock Breakout in Supernovae}}.
\bjtitle{Physical Review Letters}
\bvolume{87}(\bissue{7}),
\bfpage{071101}
(\byear{2001}).
doi:\doiurl{10.1103/PhysRevLett.87.071101}
\end{barticle}
\endbibitem

\bibitem[\protect\citeauthoryear{{Wei} and {Gao}}{2003}]{Wei+03-EpkLum}
\begin{barticle}
\bauthor{\binits{D.M.} \bsnm{{Wei}}},
\bauthor{\binits{W.H.} \bsnm{{Gao}}},
\batitle{{Are there cosmological evolution trends on gamma-ray burst
  features?}}
\bjtitle{\mnras}
\bvolume{345},
\bfpage{743}--\blpage{746}
(\byear{2003}).
doi:\doiurl{10.1046/j.1365-8711.2003.06971.x}
\end{barticle}
\endbibitem

\bibitem[\protect\citeauthoryear{{Yonetoku} et~al.}{2004}]{Yonetoku+04-EpkLum}
\begin{barticle}
\bauthor{\binits{D.} \bsnm{{Yonetoku}}},
\bauthor{\binits{T.} \bsnm{{Murakami}}},
\bauthor{\binits{T.} \bsnm{{Nakamura}}},
\bauthor{\binits{R.} \bsnm{{Yamazaki}}},
\bauthor{\binits{A.K.} \bsnm{{Inoue}}},
\bauthor{\binits{K.} \bsnm{{Ioka}}},
\batitle{{Gamma-Ray Burst Formation Rate Inferred from the Spectral Peak
  Energy-Peak Luminosity Relation}}.
\bjtitle{\apj}
\bvolume{609},
\bfpage{935}--\blpage{951}
(\byear{2004}).
doi:\doiurl{10.1086/421285}
\end{barticle}
\endbibitem

\bibitem[\protect\citeauthoryear{{Yonetoku} et~al.}{2011}]{YonEtAl:2011}
\begin{barticle}
\bauthor{\binits{D.} \bsnm{{Yonetoku}}},
\bauthor{\binits{T.} \bsnm{{Murakami}}},
\bauthor{\binits{S.} \bsnm{{Gunji}}},
\bauthor{\binits{T.} \bsnm{{Mihara}}},
\bauthor{\binits{K.} \bsnm{{Toma}}},
\bauthor{\binits{T.} \bsnm{{Sakashita}}},
\bauthor{\binits{Y.} \bsnm{{Morihara}}},
\bauthor{\binits{T.} \bsnm{{Takahashi}}},
\bauthor{\binits{N.} \bsnm{{Toukairin}}},
\bauthor{\binits{H.} \bsnm{{Fujimoto}}},
\bauthor{\binits{Y.} \bsnm{{Kodama}}},
\bauthor{\binits{S.} \bsnm{{Kubo}}},
\bauthor{\bsnm{{IKAROS Demonstration Team}}},
\batitle{{Detection of Gamma-Ray Polarization in Prompt Emission of GRB
  100826A}}.
\bjtitle{\apjl}
\bvolume{743},
\bfpage{30}
(\byear{2011}).
doi:\doiurl{10.1088/2041-8205/743/2/L30}
\end{barticle}
\endbibitem

\bibitem[\protect\citeauthoryear{{Yonetoku} et~al.}{2012}]{YonEtAl:2012}
\begin{barticle}
\bauthor{\binits{D.} \bsnm{{Yonetoku}}},
\bauthor{\binits{T.} \bsnm{{Murakami}}},
\bauthor{\binits{S.} \bsnm{{Gunji}}},
\bauthor{\binits{T.} \bsnm{{Mihara}}},
\bauthor{\binits{K.} \bsnm{{Toma}}},
\bauthor{\binits{Y.} \bsnm{{Morihara}}},
\bauthor{\binits{T.} \bsnm{{Takahashi}}},
\bauthor{\binits{Y.} \bsnm{{Wakashima}}},
\bauthor{\binits{H.} \bsnm{{Yonemochi}}},
\bauthor{\binits{T.} \bsnm{{Sakashita}}},
\bauthor{\binits{N.} \bsnm{{Toukairin}}},
\bauthor{\binits{H.} \bsnm{{Fujimoto}}},
\bauthor{\binits{Y.} \bsnm{{Kodama}}},
\batitle{{Magnetic Structures in Gamma-Ray Burst Jets Probed by Gamma-Ray
  Polarization}}.
\bjtitle{\apjl}
\bvolume{758},
\bfpage{1}
(\byear{2012}).
doi:\doiurl{10.1088/2041-8205/758/1/L1}
\end{barticle}
\endbibitem

\bibitem[\protect\citeauthoryear{{Yu} et~al.}{2015}]{Yu+15spwidth}
\begin{barticle}
\bauthor{\binits{H.-F.} \bsnm{{Yu}}},
\bauthor{\binits{H.J.} \bsnm{{van Eerten}}},
\bauthor{\binits{J.} \bsnm{{Greiner}}},
\bauthor{\binits{R.} \bsnm{{Sari}}},
\bauthor{\binits{P.} \bsnm{{Narayana Bhat}}},
\bauthor{\binits{A.} \bsnm{{von Kienlin}}},
\bauthor{\binits{W.S.} \bsnm{{Paciesas}}},
\bauthor{\binits{R.D.} \bsnm{{Preece}}},
\batitle{{The sharpness of gamma-ray burst prompt emission spectra}}.
\bjtitle{\aap}
\bvolume{583},
\bfpage{129}
(\byear{2015}).
doi:\doiurl{10.1051/0004-6361/201527015}
\end{barticle}
\endbibitem

\end{thebibliography}

\nocite{*}

\end{document}